# Physics-Informed Sequential Reconstruction of Scanning Probe Microscopy Images with Calibrated Uncertainty

Surya Prakash Reddy,[1] and Sergei V. Kalinin,[1]

[1]Department of Materials Science and Engineering, University of Tennessee, Knoxville, Tennessee 37996, USA

**Abstract**

Scanning probe microscopy (SPM) has become a mainstay of materials science, condensed matter physics, nanotechnology, biology, and semiconductor metrology. An SPM image, however, is acquired sequentially as the probe traverses the surface under closed-loop feedback while thermal drift, feedback dynamics, tip state, mechanical disturbances, and electronic noise evolve during the experiment. Image correction has therefore become an intrinsic part of SPM data analysis. Classical approaches rely on line leveling, filtering, registration, interpolation, and explicit models of tip or scanner distortions, whereas recent work increasingly uses neural networks trained to map corrupted images to corrected ones. Here, we explore a different formulation in which SPM reconstruction is treated as a Bayesian sequential state-estimation problem. The hidden state represents surface height and its low-order evolution along the slow-scan direction; an artifact-dependent quality score continuously modifies the observation likelihood; a forward Kalman pass provides the causal estimate and innovation diagnostics; and a Rauch-Tung-Striebel (RTS) backward pass incorporates information from the complete acquisition. The result is a reconstructed surface, a relative posterior uncertainty map, and line-resolved diagnostic signals. Optional components introduce lateral information and trace-retrace redundancy. For a fixed low-dimensional state the recursion is computationally lightweight and scales linearly with the number of image pixels, while the physical parameters can in principle be initialized from instrument characterization or from priors learned during previous operation, an extension not exercised here. We organize the method as a multi-axis family of models so that the contributions of state representation, artifact recognition, lateral coupling, and acquisition redundancy can be evaluated independently.

## 1. Introduction

Scanning probe microscopy (SPM) spans techniques from scanning tunneling microscopy (STM), which established atomic-scale electronic imaging,[1]to atomic force microscopy (AFM) and its many force-sensitive variants.[2] Dynamic AFM further extended this platform to a broad range of ambient and functional measurements,[3] while modern reviews emphasize that imaging performance is inseparable from scanner, feedback, and probe dynamics.[4] The characteristic aspect of all these techniques is that the image is built sequentially during a raster scan. The probe traverses the fast-scan direction, advances by one line in the slow-scan direction, and repeats this process over times ranging from seconds to many minutes. Consequently, the value recorded at each pixel reflects not only the local sample state but also the instantaneous state of the scanner, feedback loop, probe apex, environment, and acquisition electronics. The final SPM image therefore contains both sample structure and a record of how the measurement was performed.

The practical consequence is familiar to every SPM user: image correction is a routine part of the experiment. Standard software environments provide line and plane leveling, background subtraction, filtering, and interpolation as first-line tools.[5] Similar processing workflows are implemented across widely used SPM analysis packages.[6] More specialized methods remove fitted backgrounds while protecting foreground objects, correct vertical drift and illusory line slopes,[7] suppress narrow-band stripe noise in Fourier space,[8] recover sparse impulse corruption,[9] benchmark low-rank and other destriping schemes,[10] or use wavelets for high-speed denoising.[11] These operations are fast and transparent, but they are normally applied as separate post-processing steps. A Fourier filter does not know whether a removed periodicity belongs to the microscope or to the sample, and a corrected image usually carries no quantitative statement of how much of a pixel was measured and how much was inferred.

A second family of approaches uses explicit physics of image formation. Finite-tip correction treats the measured AFM image as a geometric transformation of the surface by the probe and uses blind tip reconstruction to constrain the underlying morphology.[12] Extensions of this idea address more general tip shapes and the stability of tip estimation under noise.[13] Scanner creep and hysteresis can instead be treated as deterministic positioning errors and compensated through control or calibration.[14] Feature-oriented scanning provides an acquisition-side strategy for suppressing drift and creep by repeatedly returning to surface references.[15] Thermal drift can also be estimated from additional scans or reference strips,[16] reconstructed in real space from stable lattice information,[17] or separated by combining acquisitions with different scan directions. [18] Offline tools such as unDrift make these strategies accessible for routine image correction,[19] while detector-drift compensation illustrates that even the measurement channel itself can carry a slowly evolving state.[20] These approaches are attractive for quantitative metrology because their parameters have direct physical meaning, but each typically addresses one part of the measurement chain rather than providing a common representation of uncertainty across artifacts.

Over the last several years, the most rapid growth in SPM image reconstruction has been associated with machine learning, particularly neural-network methods; the field has been reviewed in detail by Azuri and co-workers.[21] Deep networks have been used for AFM resolution enhancement,[22] scanner hysteresis compensation in hybrid model-based imaging,[23] and restoration of images acquired at increased scan speed.[24] Modern denoisers have been benchmarked against characteristic AFM noise classes,[25] whereas simulation-trained convolutional networks have been applied to atomic-resolution STM.[26] A particularly relevant example is the dual-direction ResU-Net of Kocur et al., which was trained on synthetically corrupted AFM topographies and evaluated on experimental scans.[27] Tip-related distortions have been addressed through differentiable blind tip reconstruction,[28] and through learned surface-profile recovery from virtual AFM data.[29] Recent application-specific denoisers explicitly test whether image restoration preserves morphology or mechanical analysis,[30] and super-resolution studies likewise show that image quality cannot be judged from pixel-fidelity metrics alone.[31] Collectively, these studies demonstrate that learned correction is powerful, but they also make the scarcity of clean experimental ground truth and the fidelity of the corruption model central issues.

The issue is not whether neural networks can correct SPM images—they clearly can. Rather, the question is where a learned component should enter a reconstruction problem whose acquisition order and much of whose instrument physics are known. Learned models are particularly effective when the required task is recognition of a complex probe state: autonomous STM tip conditioning is an early example.[32] Machine-learning-based tip functionalization further demonstrates that probe state can be recognized and acted upon during operation.[33] More broadly, artificial-intelligence-driven SPM has shown that learned representations can be integrated directly with instrument control.[34] These results motivate a hybrid architecture in which the acquisition physics provides the inferential backbone, while learned models are introduced where pattern recognition or an otherwise inaccessible transfer function is required.

A fourth family, drawing from both SPM and the broader inverse-problem literature, treats reconstruction probabilistically rather than as a deterministic image transform. Gaussian-process and compressed-sensing methods have already been used to reconstruct sparsely sampled SPM scans.[35] Automated microscopy frameworks have likewise emphasized the value of probabilistic decision making and explicit uncertainty.[36] Bayesian active-learning experiments in SPM provide a concrete example of this view.[37] Earlier work on large SPM data sets established the need to connect statistical representations to the physical measurement process,[38] while Bayesian model selection has been used to discriminate competing dynamical models directly from SPM spectra.[39] The mathematical foundations are provided by Kalman filtering,[40] its continuous-time and prediction extensions,[41] and fixed-interval Rauch-Tung-Striebel smoothing.[42] Standard treatments of optimal filtering[43] and Bayesian filtering and smoothing[44] provide the corresponding probabilistic framework. Adaptive covariance identification offers a route to learning measurement noise from innovations,[45] including explicitly adaptive filtering schemes.[46] Robust statistics[47] and robust state estimation[48, 49] provide complementary treatments of outliers, while jump detection[50] and change-point theory[51] formalize abrupt state changes. Two-dimensional Kalman image models were introduced decades ago by imposing a raster ordering on static images,[52] including formulations for simultaneous deblurring and denoising.[53] In SPM, by contrast, the raster order is the physical order in which the data are generated. This distinction is the starting point of the approach developed below.

Here, we formulate SPM image reconstruction as Bayesian sequential state estimation. The slow-scan coordinate is used as the recursion index, while each fast-axis position carries a low-dimensional state. A continuous quality score changes the measurement variance, a Kalman recursion provides a causal estimate, and an RTS backward pass uses later scan lines to refine earlier states. Figure 1 summarizes this acquisition-aware view. The approach is intended for intermittent observation failures and slow line-to-line evolution; it is not a substitute for tip-shape deconvolution or scan-coordinate correction.

We evaluate controlled variants along state dimension, confidence model, optional lateral regularization, pass handling, and smoothing. The primary model used for experimental interpretation is the single-pass S2-CT-G0-P1 variant (M07). Higher-order states, edge-aware lateral repair, and dual-pass fusion are treated as ablations because they do not provide uniform improvements. Validation combines synthetic ground truth, blind and known-mask baselines, held-out uncertainty calibration, bidirectional consistency checks, STM lattice preservation, and AFM calibration-grating metrology.

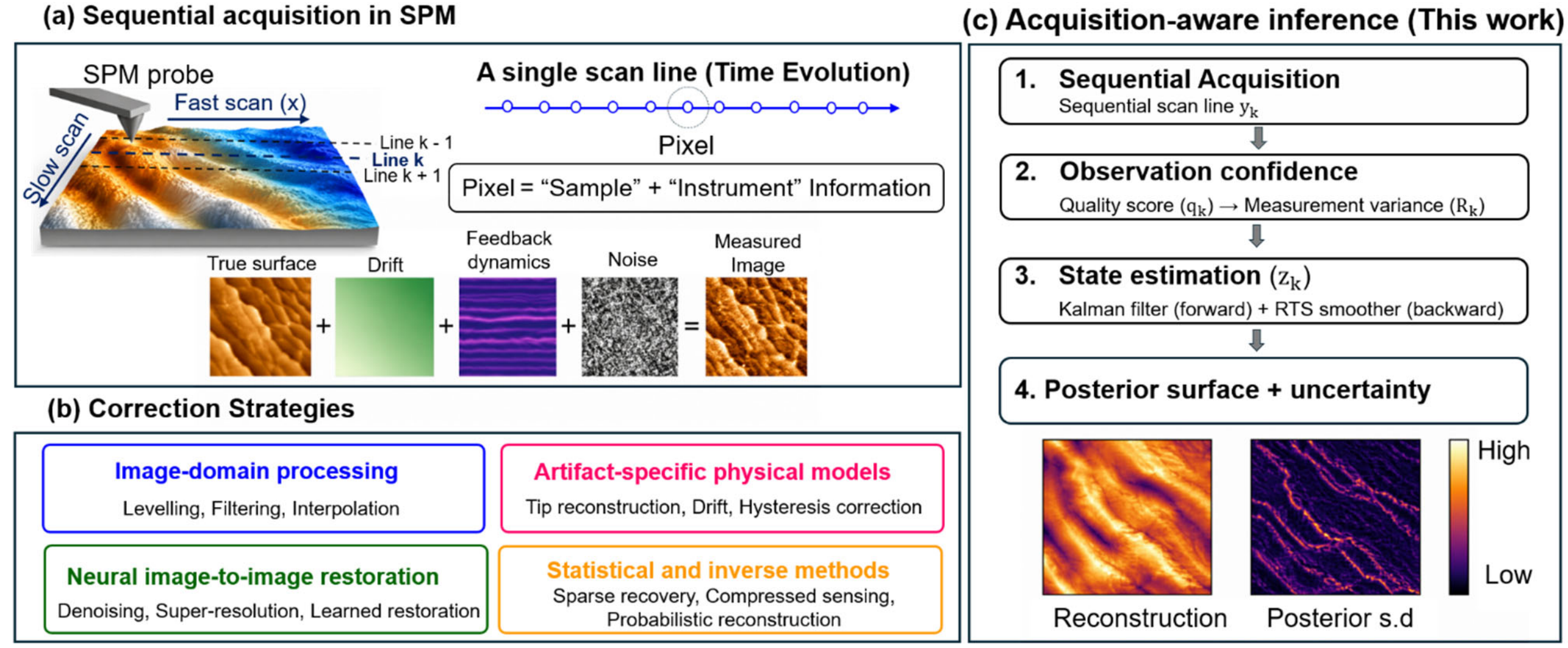

**FIG 1.** Acquisition-aware SPM reconstruction workflow from raster acquisition through quality scoring, Kalman filtering, and RTS smoothing to reconstructed height, diagnostic signals, and relative uncertainty.

## 2. Physical model and reconstruction method

The reconstruction separates three objects that are often mixed in conventional image correction: the latent state of the surface/instrument system, the observation model describing how that state is measured, and the reliability assigned to each observation. The forward filter combines these quantities in acquisition order; the backward smoother subsequently uses the completed scan to revise earlier estimates. Figure 2 summarizes this information flow.

### 2.1 SPM acquisition as a state-space process

Let x denote position along the fast-scan direction and k index successive scan lines. For each fixed x, the microscope produces a temporally ordered sequence of observations $y(k, x)$. We represent the corresponding latent state by $z(k, x)$ and write

$$\mathrm{z}_{k+1,x} = \mathbf{A}\mathrm{z}_{k,x} + \mathrm{w}_{k,x}, \qquad \mathrm{w}_{k,x} \sim \mathcal{N}(0, \mathbf{Q}), \tag{1}$$

and

$$y_{k,x} = \mathbf{C}\mathrm{z}_{k,x} + v_{k,x}, \qquad v_{k,x} \sim \mathcal{N}\left(0, R_{\mathrm{eff},k,x}\right) \tag{2}$$

In Eqs. (1) and (2), **A** represents deterministic line-to-line evolution, **Q** describes unresolved process variability, **C** maps the hidden state to the measured quantity, and $R_{\mathrm{eff}}$ describes the reliability of that measurement. For a linear-Gaussian system the filtered and fixed-interval smoothed posteriors are available analytically. [40-44] This separation is important: smooth drift or low-order surface evolution belongs in the state dynamics, unresolved variability belongs in **Q**, and temporary corruption belongs in the observation likelihood. Treating all three as additive image noise removes the physical distinction between a changing state and a failing measurement.

### 2.2 Low-order state representations

The minimal state is the constant-velocity, or S2, representation

$$\mathbf{z}_{k,x}^{(2)} = \begin{bmatrix} s_{k,x} \\ d_{k,x} \end{bmatrix}, \qquad \mathbf{A}_2 = \begin{bmatrix} 1 & 1 \\ 0 & 1 \end{bmatrix}, \qquad \mathbf{C}_2 = [1 \quad 0] \tag{3}$$

The deterministic part of Eq. (3) implies: $\hat{s}_{k+1|k,x} = \hat{s}_{k|k,x} + d_{k|k,x}$ , $d_{k+1,x} = d_{k,x}$.

Consequently, across $m$ consecutive lines whose measurements carry negligible weight, the forward predicted mean is

$$\hat{s}_{k+m|k,x} = \hat{s}_{k|k,x} + m d_{k|k,x}$$

Because the process noise in Eq. (1) is zero mean, it increases the predicted covariance through **Q** but does not alter this linear mean propagation. Here, $s_{k,x}$ is the latent height and $d_{k,x}$ is its line-to-line rate of change. The rate state necessarily combines genuine slow-axis surface slope with vertical instrument drift; with height-only measurements and no independent positional or drift reference, these contributions are not separately identifiable. Thus, when measurements are strongly down-weighted and their Kalman corrections become negligible, the forward filter traverses the resulting low-information interval using a constant-rate, linear prior. The subsequent RTS pass incorporates reliable measurements beyond the interval, constraining the final smoothed estimate from both sides.

A natural extension is to carry local gradients and curvatures as latent dynamical variables rather than estimating them by directly differentiating a noisy image. Numerical differentiation amplifies high-frequency measurement noise and becomes especially unstable where data are missing or corrupted.[54] The higher-order S6 representation therefore introduces the geometric quantities explicitly into the state and constrains their evolution through the transition matrix $\mathbf{A}_6$ and process covariance **Q**:

$$\mathbf{z}^{(6)} = [s \quad s_x \quad s_k \quad s_{xx} \quad s_{xk} \quad s_{kk}]^T, \tag{4}$$

The corresponding second-order propagator is

$$\mathbf{A}_6 = \begin{bmatrix} 1 & 0 & 1 & 0 & 0 & 1/2 \\ 0 & 1 & 0 & 0 & 1 & 0 \\ 0 & 0 & 1 & 0 & 0 & 1 \\ 0 & 0 & 0 & 1 & 0 & 0 \\ 0 & 0 & 0 & 0 & 1 & 0 \\ 0 & 0 & 0 & 0 & 0 & 1 \end{bmatrix}, \qquad \mathbf{C}_6 = [1 \quad 0 \quad 0 \quad 0 \quad 0 \quad 0]. \tag{5}$$

Only the slow-axis derivative terms directly contribute to height propagation in the present height-only observation model. The observable core $\{s, s_k, s_{kk}\}$ is a constant-acceleration state; the remaining fast-axis derivative components are not independently observable from $C = [1,0,\dots,0]$ without additional measurements or coupling assumptions. Consequently, S6 is evaluated as an exploratory ablation rather than assumed to be superior to S2.

$$\hat{s}_{k+m} = \hat{s}_k + m\hat{s}_{k,k} + \frac{m^2}{2}\hat{s}_{kk,k} \tag{6}$$

Physically, $s$ is the latent height, $s_x$ and $s_k$ are local fast- and slow-axis slopes, and the remaining terms represent local curvature. Because the current observation contains height only, these labels describe the transition model rather than separately identified physical fields. State clipping used for numerical robustness is documented in the Supplementary Material.

The remaining state variability is represented by the process covariance

$$\mathbf{Q} = diag(\sigma_1^2, \dots, \sigma_n^2), \quad (7)$$

where n is the state dimension ($n = \dim(z)$)

The entries of Q are not only fitting parameters. They may be estimated from clean regions, repeated scans, or instrument-characterization data after removal of deterministic line-to-line trends. The transition matrix A fixes the assumed kinematics, whereas Q sets the allowed departure of the microscope/sample system from that kinematics. The Supplementary Material therefore distinguishes fixed structural parameters, tunable covariances, automatically estimated diagnostics, and derived quantities; persistent instrument priors are presented only as a prospective extension.

**2.3 Observation reliability as a continuous likelihood weight**

Artifact handling enters through the observation model. Each measurement is assigned a quality $q(k,x)$ in $(0{,}1]$, and the baseline measurement variance is inflated according to ($\varepsilon = 10^{-6}$ is a numerical floor)

$$R_{eff,k,x} = \frac{R_k}{max(q_{k,x}, \varepsilon)} \quad (8)$$

For $q = 1$, the observation is used normally. As q approaches zero, $R_{\mathrm{eff}}$ grows, the Kalman gain approaches zero, and the recursion increasingly relies on the state prediction. This is a soft form of robust estimation: partially reliable data can still contribute rather than being hard-rejected.[47, 49, 55] Importantly, q is external to the state recursion. It can be supplied by an interpretable detector, a learned classifier, an auxiliary instrument channel, or a human annotation without changing Eqs. (1)-(2).

In the reference implementation, quality is computed once per scan line from four robust statistics: deviation from the line median ($q_{crash}$), excess high-frequency content measured as the standard deviation of adjacent-sample differences ($q_{osc}$), disagreement with the mean of the previous five lines ($q_{cons}$), and a tip-change score formed from the ratio of the line innovation RMS to its post-burn-in reference ($q_{tip}$); in the dual-pass P2 variants the consistency factor additionally uses trace-retrace disagreement. Their bounded product gives $q_k$, which is broadcast across that line. The notation $q(k,x)$ allows future pixelwise detectors, but the results reported here use linewise $q_k$.

$$q_k = q_{\mathrm{crash},k}\, q_{\mathrm{osc},k}\, q_{\mathrm{cons},k}\, q_{\mathrm{tip},k} \quad (9)$$

The detector is intentionally interpretable, but its thresholds are scale dependent. It is effective for line-wide crashes and some oscillatory failures; it cannot reliably localize short within-line defects and can misclassify periodic or step-rich surfaces. These limitations motivate the artifact-stratified analysis and the calibration-grating stress test presented below.

### 2.4 Forward Kalman recursion and adaptive measurement noise

For each fast-axis position x, the forward pass processes scan lines in acquisition order. The prior mean and covariance are

$$\hat{\mathbf{z}}_{k|k-1} = \mathbf{A}\hat{\mathbf{z}}_{k-1|k-1}, \tag{10}$$

$$\mathbf{P}_{k|k-1} = \mathbf{A}\mathbf{P}_{k-1|k-1}\mathbf{A}^T + \mathbf{Q} \tag{11}$$

The innovation, its variance, and the Kalman gain are

$$\nu_k = y_k - \mathbf{C}\hat{\mathbf{z}}_{k|k-1}, \qquad S_k = \mathbf{C}\mathbf{P}_{k|k-1}\mathbf{C}^T + R_{\mathrm{eff},k}, \tag{12}$$

$$\mathbf{K}_k = \mathbf{P}_{k|k-1}\mathbf{C}^T S_k^{-1} \tag{13}$$

The posterior state and covariance are then

$$\hat{\mathbf{z}}_{k|k} = \hat{\mathbf{z}}_{k|k-1} + \mathbf{K}_k \nu_k, \tag{14}$$

$$\mathbf{P}_{k|k} = (\mathbf{I} - \mathbf{K}_k\mathbf{C})\mathbf{P}_{k|k-1} \tag{15}$$

The innovation in Eq. (12) is more than an internal residual: it is a line-resolved diagnostic generated by the estimator at every acquisition step. It records how strongly the new observation disagrees with the one-step prediction. Abrupt changes in innovation energy are the natural statistic for detecting jumps in dynamical systems,[50] while sustained changes are a standard signal of model or covariance mismatch.[45, 46] In SPM, a sudden increase can therefore flag a crash or tip-state transition, periodic structure in the residual can accompany feedback ringing, and persistent excess variance can indicate that the assumed measurement noise or process model is no longer adequate. The causal estimator defined by Eqs. (10)-(15) therefore provides both reconstruction and instrument-facing diagnostics.

The baseline measurement variance R may be updated after burn-in by innovation-covariance matching, subject to a high-quality-line gate and a floor $R_{min} = 0.1R_0$ ($\alpha_R = 0.95$; burn-in = $\max(25, 0.1N_k)$ lines). In the present real data runs the estimate is nearly constant, so adaptive R is treated as a monitoring safeguard rather than a demonstrated source of performance gain.

$$\hat{R}_k = \alpha_R \hat{R}_{k-1} + (1 - \alpha_R) max\left(\langle \nu_k^2 \rangle_x - \boldsymbol{C}\langle \boldsymbol{P}_{k|k-1} \rangle_x \boldsymbol{C}^T, R_{min}\right), \tag{16}$$

Equation (16) is evaluated only after the initial transient and only when mean line quality exceeds 0.7. The gate prevents a transient crash from being absorbed into the baseline noise model. The implementation details and parameter values are provided in the Supplementary Material.

### 2.5 Non-causal reconstruction with the RTS smoother

The forward Kalman filter is causal: the estimate at line k is conditioned only on measurements available up to k. For offline reconstruction, however, the entire image is already available. The Rauch-Tung-Striebel (RTS) fixed-interval smoother uses the stored filtered and predicted state distributions from the forward pass and then performs a second recursion from the final line back to the beginning.[42-44] At each step the smoother gain determines how strongly information from later observations should revise the earlier filtered state.

The backward recursion is

$$\mathbf{G}_k = \mathbf{P}_{k|k}\mathbf{A}^T\left(\mathbf{P}_{k+1|k}\right)^{-1}, \tag{17}$$

$$\hat{\boldsymbol{z}}_{k|N} = \hat{\boldsymbol{z}}_{k|k} + \boldsymbol{G}_k\left(\hat{\boldsymbol{z}}_{k+1|N} - \hat{\boldsymbol{z}}_{k+1|k}\right), \tag{18}$$

$$\boldsymbol{P}_{k|N} = \boldsymbol{P}_{k|k} + \boldsymbol{G}_k\left(\boldsymbol{P}_{k+1|N} - \boldsymbol{P}_{k+1|k}\right)\boldsymbol{G}_k^T. \tag{19}$$

The value of the backward pass is most transparent across a sequence of unusable lines. The forward filter enters the gap with the last reliable state and propagates open loop while its covariance grows. The backward pass then introduces the first reliable post-gap measurements and propagates their information toward earlier lines using Eqs. (17)-(19). The final reconstruction is constrained from both sides of the gap. With S2 model, this gives an uncertainty-weighted linear bridge; with the observable second-order slow-axis state it gives a parabolic bridge whose curvature is inferred from the surrounding acquisition.

The reconstructed surface and its posterior standard deviation are

$$\hat{s}_{k,x} = \left[\hat{\boldsymbol{z}}_{k|N,x}\right]_1, \qquad \sigma_{k,x} = \sqrt{\left[\boldsymbol{P}_{k|N,x}\right]_{11}}. \tag{20}$$

The square root of the smoothed height variance in Eq. (20) is reported as a relative uncertainty diagnostic. It grows during low-information intervals and ranks pixelwise error on held-out synthetic data, but clipping, heuristic quality weights, optional post-update smoothing, and model mismatch invalidate an unqualified Gaussian-posterior interpretation. Quantitative intervals therefore use an empirical calibration fit on a disjoint validation set.

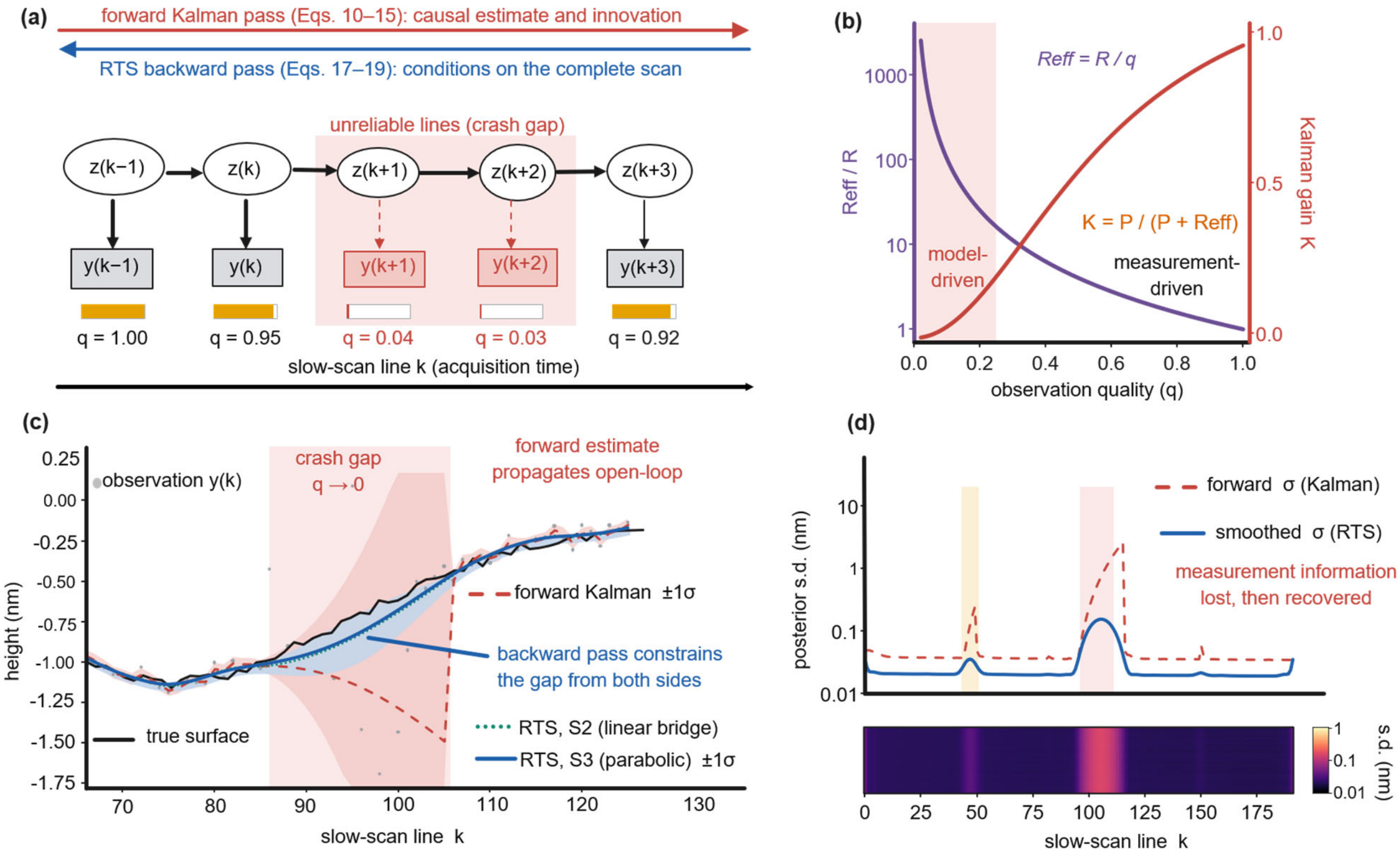

**FIG 2**. Mathematical anatomy of the reconstruction. (a) State-space chain. (b) Quality-dependent Kalman gain. (c) Forward and RTS estimates across a synthetic crash. (d) Relative uncertainty across the gap. $R_{eff} = \frac{R}{\max(q,\varepsilon)}$

### 2.6 Optional lateral and dual-pass information

The primary M07 estimator treats fast-axis columns independently and propagates state information along the slow-scan direction. Neighboring fast-axis samples may provide additional constraints where individual estimates are uncertain. We therefore evaluate an optional edge-aware lateral regularizer that weights neighboring height estimates by lateral distance, an edge barrier, measurement quality, and inverse estimated variance. The update is applied after the forward Kalman pass and only where the estimated height uncertainty exceeds a prescribed threshold. Although this construction is motivated by bilateral filtering and edge-preserving diffusion, it modifies the filtered heights without consistently propagating the resulting cross-pixel covariance through the RTS recursion. It is therefore treated as heuristic lateral regularization and evaluated as an ablation, rather than interpreted as a Gaussian-process posterior.

Trace and retrace provide a second potential source of information. After orientation correction and registration, the two acquisitions sample approximately the same surface locations at different times and may contain different realizations of measurement noise and transient failure. The implemented P2 variant classifies the two directions independently and forms a quality-weighted observation,

$$w_f = \frac{q_f}{q_f+q_b+\varepsilon},\ w_b = 1 - w_f,\ y_{\text{fused}} = w_f y_f + w_b y_b. \tag{21}$$

The fused observation is assigned quality $q = \max(q_f, q_b)$ and baseline measurement variance $\frac{R}{2}$ before quality inflation. Under independent measurement errors, equal baseline noise, and a common latent height, the quality-weighted mean agrees with information-form fusion when numerical safeguards are neglected. However, the corresponding information-form variance is $\frac{R}{q_f+q_b}$, whereas the implemented rule assigns $\frac{R}{2\max(q_f,q_b)}$. These variances agree when the quality scores are equal; when they differ, the implemented rule underestimates measurement uncertainty under the stated assumptions. Registration alone does not establish error independence or eliminate state evolution between the two acquisitions.

The implemented P2 variant increases average reconstruction error in the factorial benchmark and is therefore retained as an exploratory comparison rather than included in the primary M07 estimator. Detailed component effects are reported in Section 3.2. Experimental opposite-pass measurements are used chiefly as held-out consistency references, not as clean ground truth. These findings concern the present fusion rule and do not establish that dual-pass information is intrinsically detrimental. A probabilistically consistent extension would combine the two measurement likelihoods using their individual effective variances and account for cross-pass error correlations or temporal state evolution where necessary.

### 2.7 Implementation and validation protocol

All reported model variants share the same normalization, initialization, and core recursion. Synthetic cases are generated from smooth sinusoidal, step-terrace, and power-law rough surfaces and corrupted by clean, crash, drift, oscillation, tip-change, or combined conditions. The full factorial archive contains 2880 reconstruction runs. External baseline hyperparameters are selected on ten separately seeded tuning images, and uncertainty calibration uses 54 cases disjoint from 54 held-out test cases. Real-data analyses use independently acquired opposite passes, crystalline reciprocal-space peaks, or nominal calibration geometry as reference information rather than treating an unprocessed image as ground truth. Complete parameters, splits, and file locations are given in the Supplementary Material.

## 3. Results and Discussion

Validation proceeds from controlled synthetic tests to experimental checks without clean pixel-level ground truth. Synthetic tests quantify reconstruction error, spectral fidelity, collateral modification, and interval coverage. Experimental tests examine scan-reversal consistency, retention of STM lattice peaks, and AFM calibration-grating metrology. Conclusions are tied to the reference available in each setting.

### 3.1 Reconstruction of controlled artifact classes

Figure 3 shows four crash configurations drawn from the controlled benchmark: a single failed line, gaps of 8 and 32 lines, and simultaneous failures in both passes. The reconstruction bridges missing intervals when the state model remains predictive, while relative uncertainty increases toward the least constrained part of a gap. These examples test intermittent observation loss; drift, oscillation, and tip-change strata enter the factorial analysis but are not represented by the image montage.

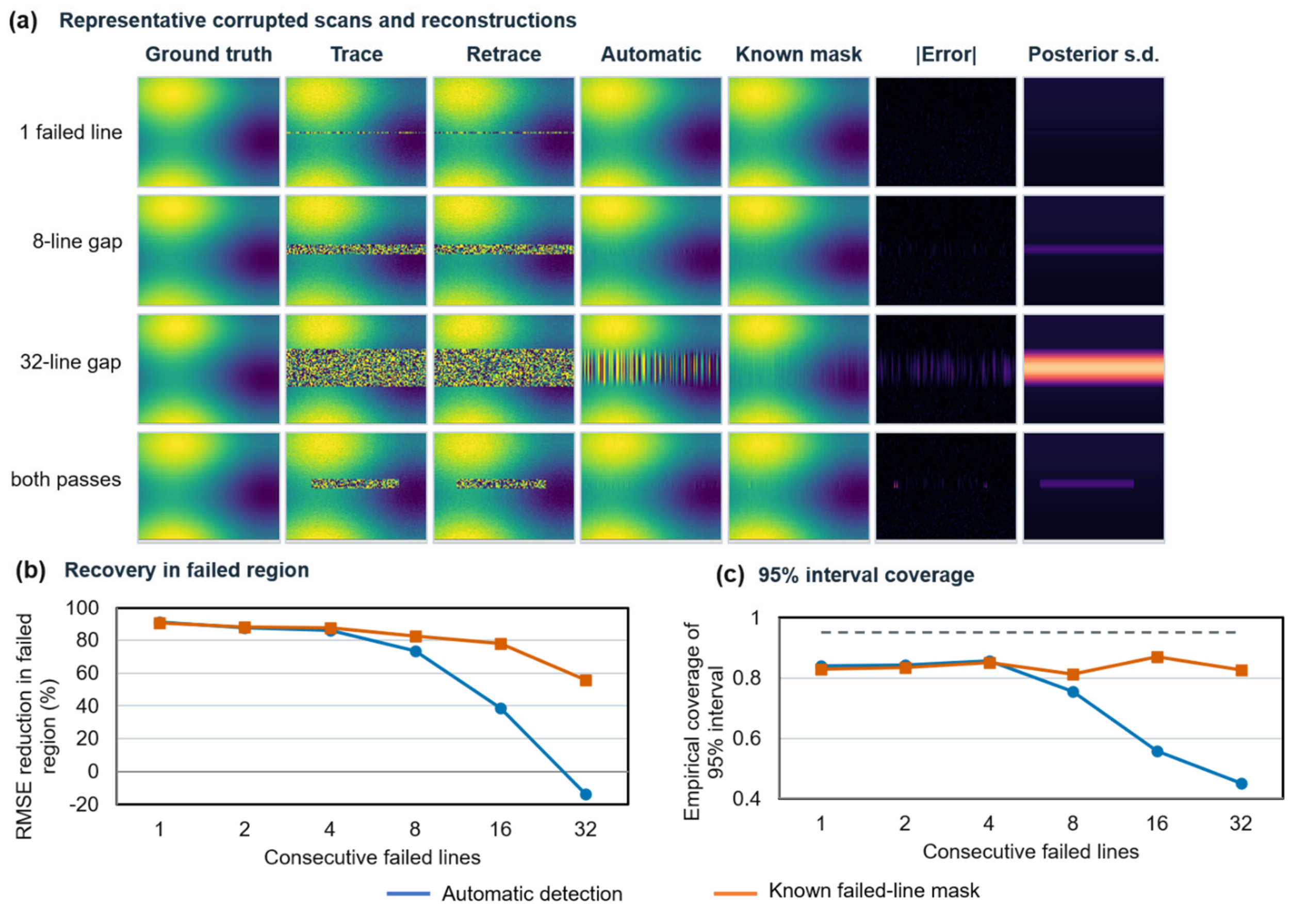

**FIG 3.** Controlled synthetic crash reconstructions. Rows show a one-line failure, 8-line and 32-line gaps, and both-pass corruption. Columns show ground truth, corrupted observations, inferred failure regions, reconstruction error, reconstructed surface, relative uncertainty, and line profiles.

### 3.2 Controlled model comparisons

The model family asks which information is useful when scan lines become unreliable. The low-order state represents height and its line-to-line rate; the higher-order state allows curvature but introduces quantities that height-only measurements cannot independently identify. Confidence weighting changes how strongly a measurement influences the estimate. Lateral regularization draws on neighboring columns, while pass fusion combines trace and retrace. RTS smoothing addresses a different question: whether reliable observations acquired after a damaged interval help constrain its reconstruction.

The paired comparisons favor reliable measurements and temporal context over added model complexity (Fig. 4). Confidence weighting improves pooled reconstruction, with its clearest benefit for crashes, because unreliable lines contribute less to the observation likelihood. RTS smoothing improves recovery by bringing information from the far side of a gap into the estimate. Neither benefit is universal: the detector can down-weight legitimate structure, and interpolation becomes unreliable when a missing interval exceeds the scale over which the state model remains predictive.

Figure 4 reports paired percentage changes in RMSE, with all other model axes held fixed. Positive values mean that switching the tested component reduced RMSE; negative values mean that it increased RMSE. Means and 95% intervals were obtained from 10,000 case-level bootstrap draws. The pooled comparisons contain 480 paired cases for quality weighting and lateral regularization, 720 for the higher-order state and pass fusion, and 1440 for RTS smoothing. The horizontal scale is linear within ±5% and logarithmic outside that range so that the large pass-fusion effect and the smaller component effects remain visible together. The full archive contains 120 synthetic cases, 12 implemented variants, and both forward and RTS outputs, giving 2880 model-stage records.

Higher-order states have no resolved average advantage in this experiment. The lateral regularizer can alter real texture, and the implemented pass-fusion rule worsens the pooled error. These findings support M07—the low-order, quality-weighted, single-pass model with RTS smoothing—as the primary model for experimental interpretation. They do not imply that lateral information or a second pass is intrinsically unhelpful; the present operators and their uncertainty treatment require improvement. The radar comparison in Fig. 8 shows how these trade-offs depend on corruption regime. Exact paired effects and model-by-model results are retained in Table S2, Fig. S2, and Sec. S15.

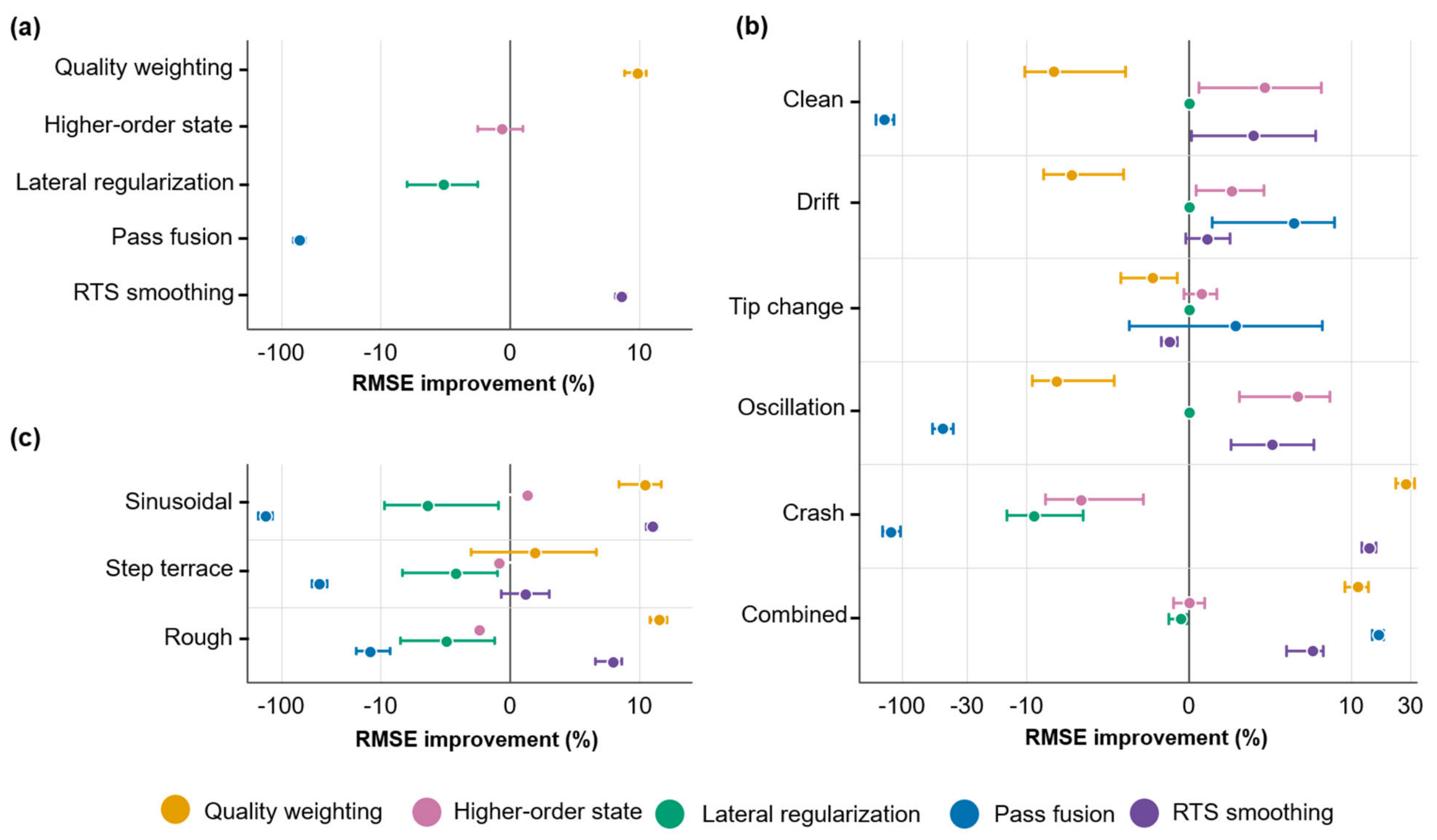

**FIG 4**. Paired component effects on RMSE. (a) Pooled benchmark. (b) Artifact classes. (c) Surface families. Markers show mean percentage change and horizontal bars show 95% bootstrap intervals; positive values indicate improvement. Colors identify component contrasts.

### 3.3 Metrics for physically meaningful reconstruction

A useful reconstruction must recover damaged regions without erasing undamaged structure. We therefore evaluate local error alongside structural similarity, spectral content, roughness, and changes in clean regions. Uncertainty has a separate role: a map can correctly rank difficult pixels while still assigning intervals that are too narrow. In the held-out calibration experiment, raw intervals under-cover, and empirical calibration brings coverage closer to its nominal level without reaching it exactly (Fig. 5E). This experiment uses the archived M08 configuration and demonstrates the calibration procedure; it does not establish calibrated uncertainty for M07 or transferability to new instruments. Calibration splits, full coverage values, and the separate GP comparison are given in Sec. S7, Table S4, and Fig. S3.

### 3.4 Experimental data and external baselines

The baseline comparison separates recognition of a failed measurement from reconstruction of the missing surface (Fig. 5). Total variation favours piecewise smooth structure, non-local means averages similar image patches, compressed sensing uses a sparse representation, and Gaussian-process reconstruction imposes spatial correlation. When failure locations are unknown, the archived sequential workflow achieves lower mean failed-region error than these baselines, but its advantage diminishes as the gap widens. Once the true mask is supplied, classical methods become competitive or better. The main benefit is therefore identifying unreliable lines and adjusting their likelihood weight, rather than a uniquely superior interpolation rule. This archive uses M10, including exploratory lateral regularization and pass fusion, so its numerical ranking must not be transferred to primary M07.

All synthetic surfaces used in Figure 5 are normalized to the 0–1 range before corruption. Panel A therefore uses a common 0–1 height scale and a common 0–0.30 absolute-error scale; the ground-truth column has no error tile because its error is identically zero. Panels B–D average 30 cases spanning five surface families and six gap widths. Baseline hyperparameters were selected using ten separately seeded tuning images that were not included in the benchmark. In panel C, the 0.01 clean-region threshold is the tolerance used during that selection. The shaded region marks values above this threshold. In panel D, the horizontal references denote raw-input spectral error and zero roughness bias; the archived Kalman series has no roughness-bias value. In panel E, the diagonal denotes nominal coverage, so points below it are over-confident.

The archived sequential comparator in Figure 5B–D is M10 (S2, CT, GC, P2), whereas the held-out calibration procedure in Figure 5E uses M08 (S2, CT, G0, P2). The two panels use distinct evaluation sets and are not pooled. They exercise the benchmark workflow rather than provide a direct validation of primary M07. The separate comparison with Gaussian-process interval estimates uses 12 crash-gap cases and remains in Fig. S3. The figure panels and their plot data are generated from the archived tables by the accompanying Python script and notebook.

Recovery also carries a cost. Blind sequential reconstruction changes some clean regions and has poorer spectral fidelity than the input in this benchmark. Supplying the true mask greatly reduces the spectral penalty, implicating failure detection as an important source of unwanted modification. A visually smoother result should therefore be accepted only when the target features and downstream measurements are preserved. All gap-width results, clean-region changes, and spectral metrics remain in Sec. S6, Table S3, and Sec. S15.

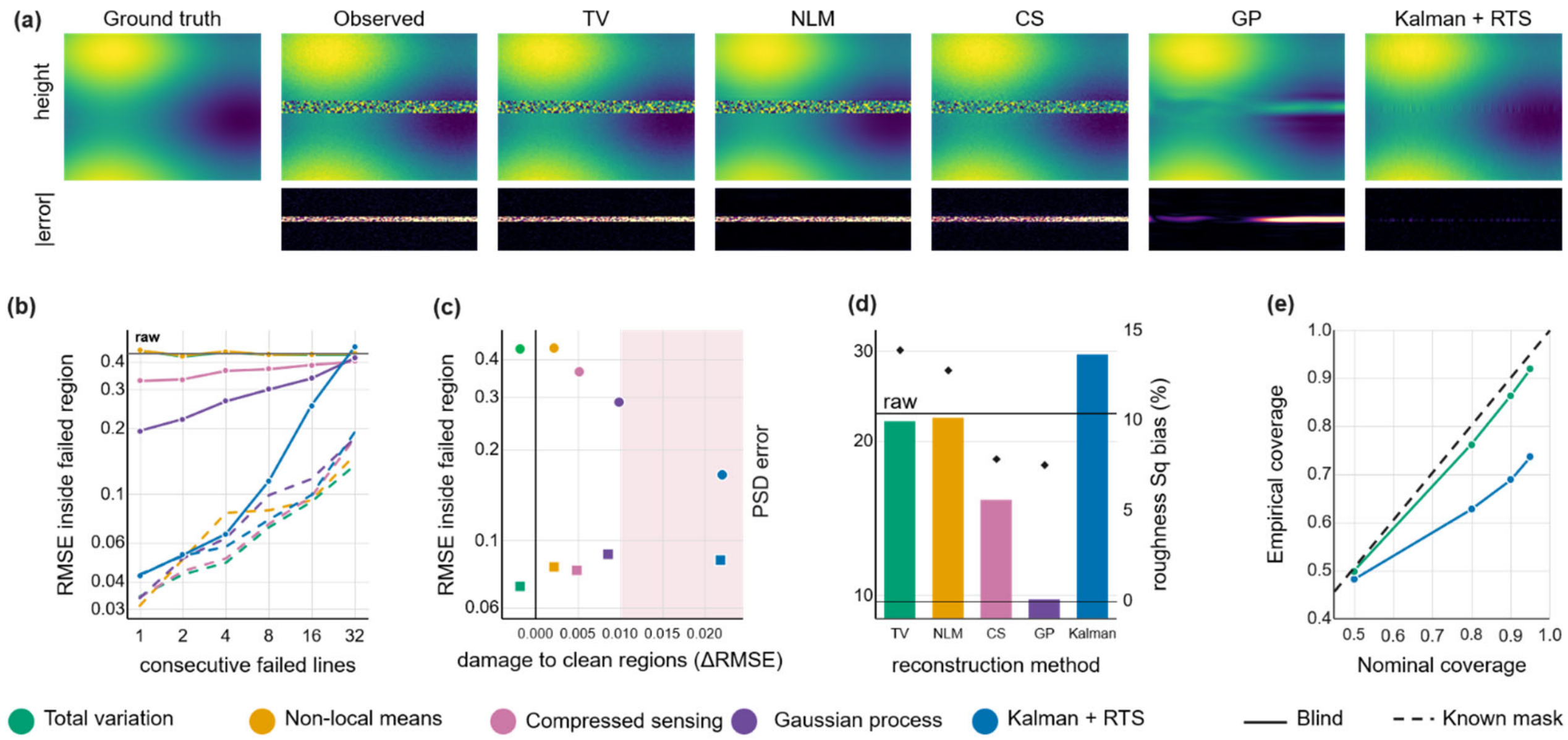

**FIG 5.** External baselines and interval calibration. (a) Blind recovery of an eight-line synthetic crash, with absolute error below. (b) Failed-region RMSE versus gap width. (c) Failed-region recovery versus clean-region modification. (d) Spectral error and roughness bias. (E) Empirical interval coverage. Colors identify methods, while line and marker styles distinguish blind and known-mask reconstruction.

Red3 is a 256 × 256-pixel AFM height acquisition over a 10 μm field. For Figure 6, each pass is reconstructed independently after percentile normalization and then mapped back to its input height units. The panels use a shared height scale and are shown in their stored coordinates without an additional reversal or registration. The orientation audit is ambiguous, and the registered correlation is approximately 0.33. The comparison therefore illustrates pass-specific estimator behavior and does not establish registered agreement or absolute accuracy.

An opposite acquisition direction can provide a consistency reference, but it contains its own artifacts and may not share an unambiguous spatial alignment. Figure 6 reconstructs the two Red3 passes independently and displays source-specific quality. The stored audit marks this acquisition's orientation as ambiguous, so the image comparison is a diagnostic example rather than evidence of registered agreement. More generally, consistency depends on registration and on whether the reference is processed in the same way; it cannot substitute for clean-ground-truth accuracy. The paired-data audit and comparison definitions are retained in Sec. S8.

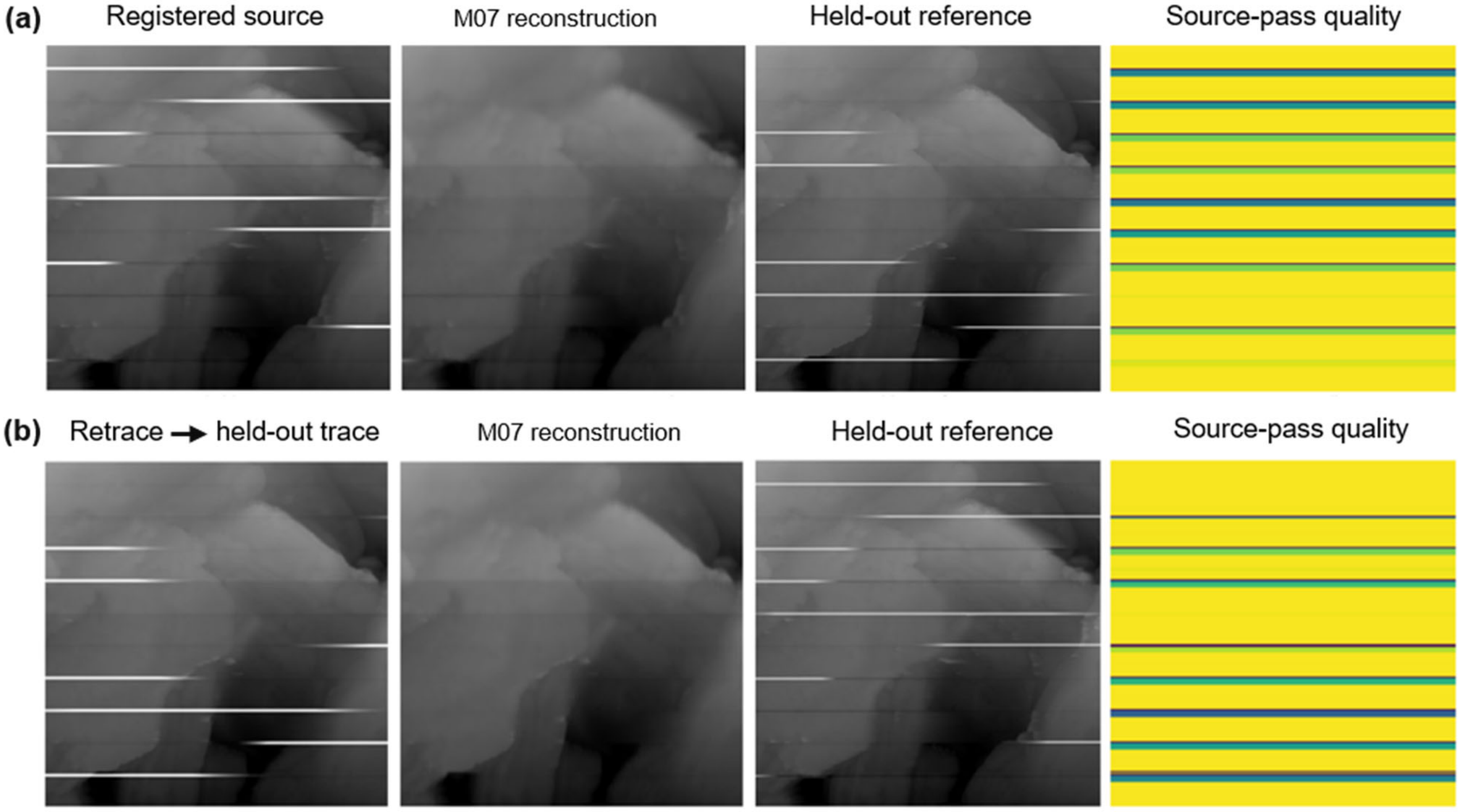

**FIG 6**. Single-pass M07 diagnostics for Red3 AFM height data. (a) Trace as the source. (b) Retrace as the source. Columns show the source, reconstruction, opposite pass, and source-only quality. Height is in nanometers and quality is dimensionless (0–1).

### 3.5 Physical validation without pixel-level ground truth

For atomic-resolution STM, preservation of lattice information provides a physical check beyond visual smoothness (Fig. 7). Successful artifact suppression should reduce scan-line contamination while retaining the position and amplitude of reciprocal-space peaks. The single-pass Kalman variants show this balance across the six acquisitions, whereas stronger smoothing can suppress lattice peaks together with the background. Agreement with an accepted period is claimed only where the surface assignment is unambiguous. The example now displays primary M07; quantitative panels retain the full method comparison. Acquisition conditions, spectral definitions, exact retention and suppression values, and unassigned periods are documented in Sec. S10 and Sec. S15.

The Figure 7 example is Si100-Ge-3, identified in the supplied metadata as Si(100)-2×1-Ge and acquired over 12 × 12 $nm^2$ at −1.9 V sample bias and 0.1 nA tunnelling current with 400 × 400 pixels; Ge concentration is not recorded. The real-space panels share the raw image's 1st–99th-percentile intensity limits, while each Fourier magnitude uses its own 40th–99.99th-percentile logarithmic limits over ±3 $nm^{-1}$. Suppression is defined as the raw-to-reconstructed ratio. Panel C pools the six acquisitions across either the six single-pass Kalman variants or the four classical baselines. In panel D, STM 1–4 denote Si100-Ge-1, Si111-1, Si100-Ge-2, and Si100-Ge-3. The two Au/Si acquisitions are excluded because their dominant periods lack an unambiguous assignment. These reciprocal-space checks evaluate structural preservation, not clean pixelwise accuracy.

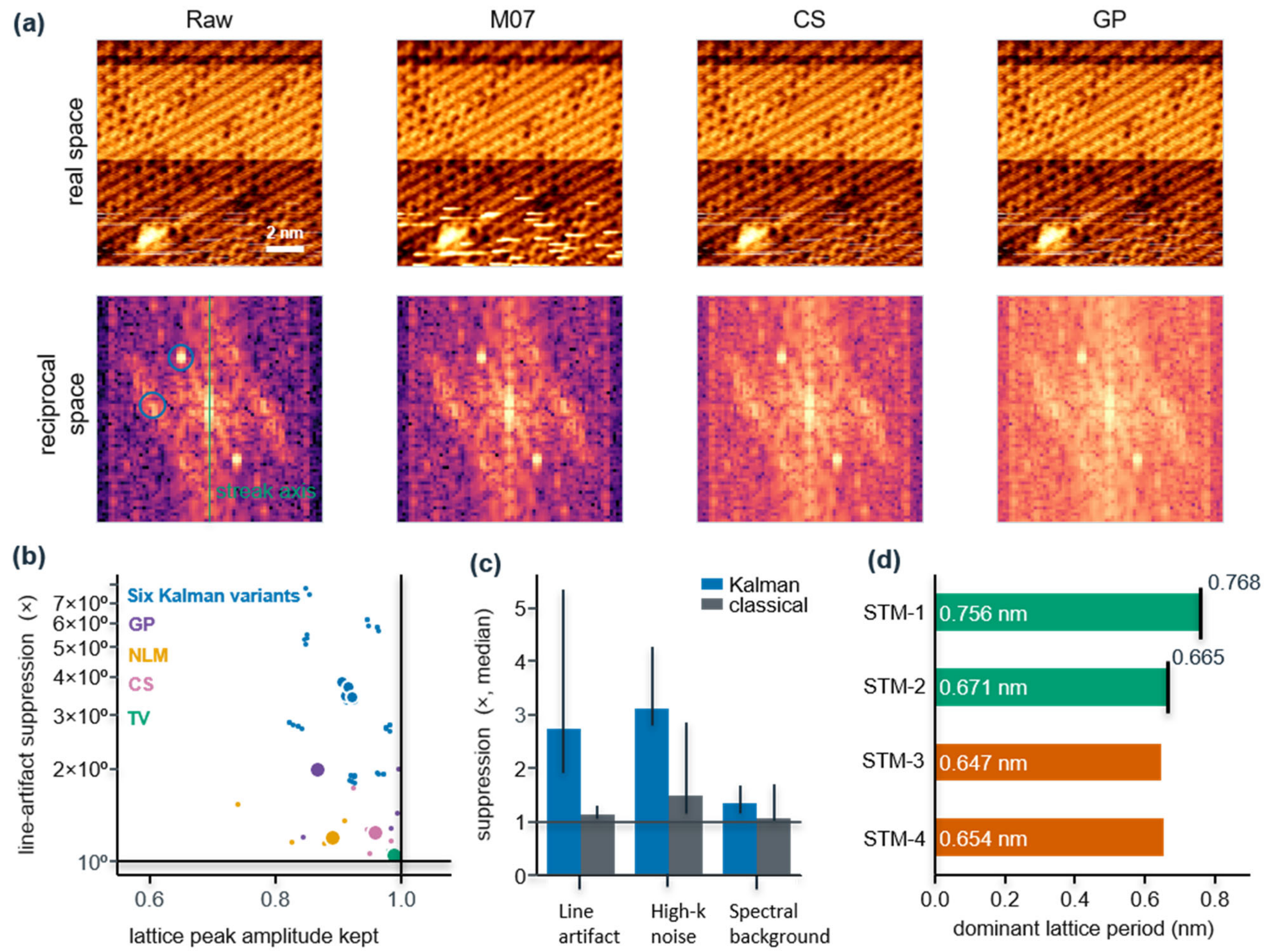

**FIG 7.** Lattice preservation in real STM data. (a) Real-space images and Fourier magnitudes for the raw input, M07, compressed sensing (CS), and Gaussian-process reconstruction (GP). (b) Lattice-peak retention versus line-artifact suppression; faint points are acquisitions and solid points are method means. (c) Median suppression with interquartile ranges. (d) Dominant periods; blue and grey denote assigned and unassigned periods, and black ticks show reference values.

### 3.6 AFM calibration-grating metrology and detector stress test

The calibration grating tests whether image correction preserves dimensions that would be reported in a metrology workflow. Most variants retain the pitch and step height relative to the supplied raw scans, while the unweighted dual-pass control introduces a substantial step-height bias. Confidence weighting improves repeatability in this small set, but the detector assigns very low quality to one strongly periodic scan. Here, legitimate sample structure resembles the statistics of an artifact.

This failure sets a practical boundary for automated correction: detector thresholds must be checked against the length and intensity scales of the sample. The available grating metadata specify nominal dimensions but provide no certified step-height uncertainty, so this experiment establishes geometric self-consistency rather than absolute metrological accuracy. The complete A–D grating figure has been moved to Fig. S7, with method-level results in Table S6 and Sec. S15.

Figure 8 summarizes the 120-case factorial benchmark over all 12 variants after RTS smoothing. For each metric and corruption regime, the lowest and highest mean values across the variants define scores of 0 and 100. Error metrics are reversed so that larger scores always indicate better relative performance. The scores are valid only for comparing variants within one spoke; they are not absolute errors, effect sizes, or uncertainty estimates, and polygon area is not a performance statistic. Unnormalized metrics and runtimes are reported in Fig. S2 and the distributed tables. The AFM calibration-grating results previously presented as Figure 8 are now provided as Fig. S7.

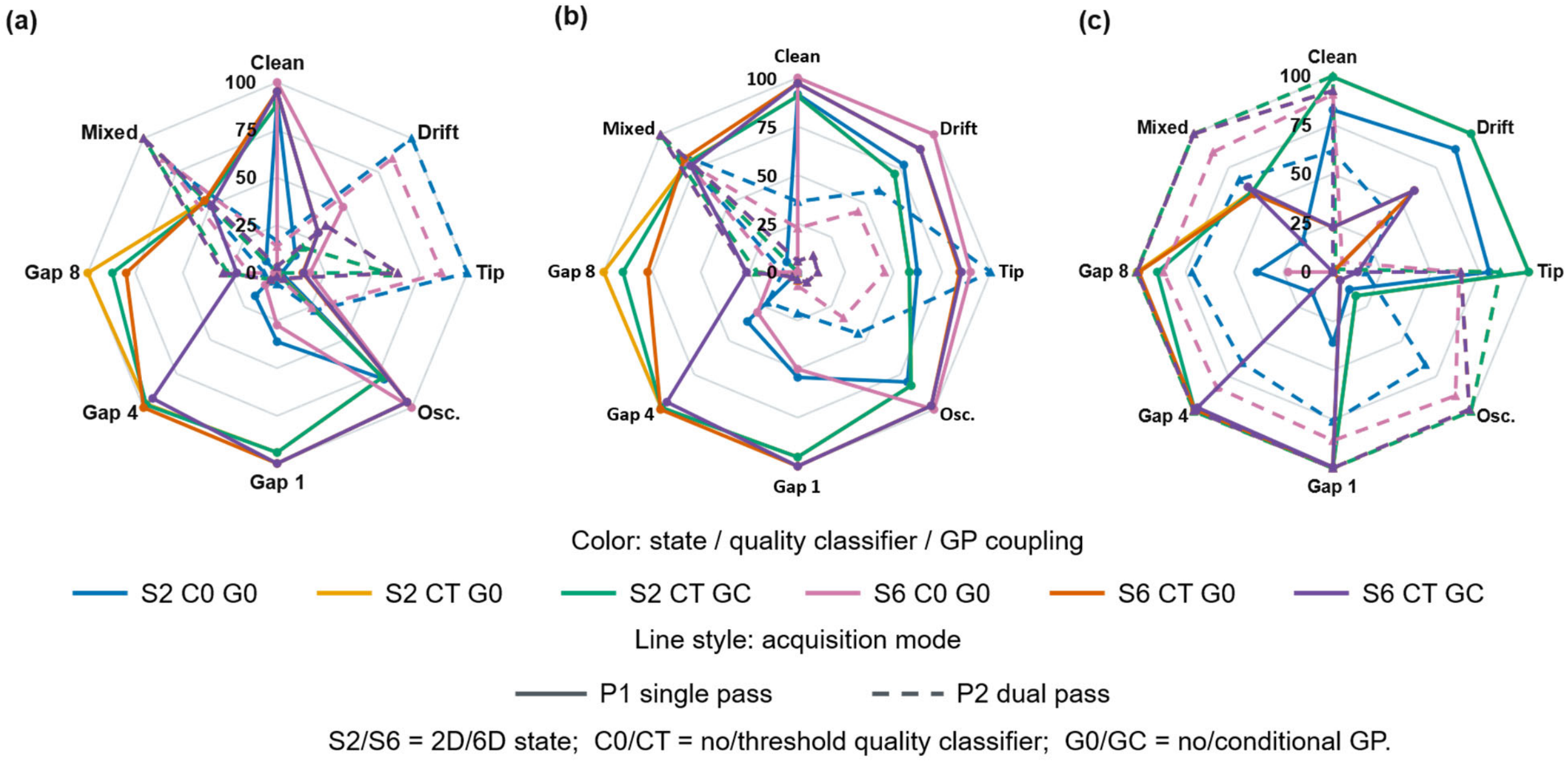

**FIG 8.** Artifact-resolved model-family comparison after RTS smoothing. (a) RMSE. (b) SSIM. (c) Radial power-spectral-density error. Each spoke is a corruption regime; larger normalized scores indicate better relative performance. Colors identify model families, and solid and dashed lines identify P1 and P2.

The physical value of sequential estimation is that it distinguishes a changing apparent-height field from an unreliable observation of that field. A low-order trend supplies a short-range prediction, quality weighting limits the effect of damaged lines, and smoothing brings in measurements from the far side of a gap. The synthetic experiments support this mechanism most clearly for intermittent line failures. They also show its limit: as the missing interval grows, reconstruction becomes dominated by assumptions about the surface rather than information in the measurements.

The ablations favor M07 as a transparent primary configuration, but do not establish universal superiority. Quality weighting can reject genuine structure; higher state dimension does not resolve unobservable components; lateral regularization alters texture without a complete covariance update; and the present dual-pass likelihood is overconfident for unequal pass qualities. The M10 baseline archive and M08 calibration archive answer useful methodological questions, but a matched baseline and calibration study of M07 is still needed to attach those numerical claims to the primary model.

Uncertainty should be interpreted at two levels. The propagated covariance indicates where information has been lost under the assumed model. Empirical coverage tests whether intervals based on that covariance are appropriate for a defined population of surfaces and artifacts. Calibration improves coverage in the held-out M08 experiment but leaves residual under coverage. It cannot repair an incorrect surface model, recover an unobserved feature, or guarantee transfer across operating conditions.

Experimental evidence is correspondingly specific. Independent passes can reveal direction-dependent behavior, but ambiguous registration limits agreement claims. Lattice peaks test preservation of periodic structure, and gratings test dimensional consistency; neither supplies complete pixelwise truth. These checks connect image processing to the scientific quantity of interest and prevent visual smoothness from becoming the sole acceptance criterion.

The current state model does not explicitly identify finite-tip broadening, lateral coordinate errors, hysteresis, or detailed feedback dynamics. Those mechanisms require additional observation or coordinate models and independent constraints. Useful extensions include sample-aware pixelwise reliability, covariance-consistent lateral coupling, and a joint model of multiple passes. Their benefit should be established using both artifact recovery and preservation of undamaged physical observables.

## 4. Conclusion and future work

SPM reconstruction can be organized around the way the measurement is acquired as a filtering-reconstruction problem. A height/rate state predicts local apparent-height evolution, linewise quality controls the influence of each observation, and RTS smoothing constrains damaged intervals using reliable lines on both sides. In the controlled model family, quality weighting and smoothing reduce pooled RMSE by 9.5% and 6.6%; more complex states and the implemented lateral and dual-pass operations provide no uniform benefit.

The resulting framework is found to be most effective for intermittent line failures over a limited predictive range. STM lattice checks and AFM dimensional tests demonstrate useful preservation criteria and reveal detector sensitivity to actual physical structure. Reconstructed height should therefore be interpreted together with its acquisition diagnostics, relative uncertainty, and the physical observable being measured.

A nonlinear observation model could be paired with unscented filtering,[66] provided that its additional parameters are identifiable from the available measurements. Real-time deployment would also require acquisition and computing interfaces; edge-assisted electron microscopy offers an adjacent implementation precedent,[67] rather than validation of the present SPM workflow. Probabilistic digital twins provide a framework for updating a physical model from observations,[68] and Bayesian conavigation connects such updating with active materials exploration.[69] A recent preprint explicitly couples sample and instrument twins for predictive microscopy.[70] These studies motivate future models that distinguish sample evolution from instrument response.

## Supplementary Material

See the supplementary material for model-variant definitions, implementation and state-clipping details, quality-detector features and thresholds, benchmark construction and data splits, baseline hyperparameters, uncertainty calibration, paired-pass consistency analysis, STM and AFM analysis procedures, calibration-grating results, exploratory dual-pass diagnostics, and additional model details.

## Acknowledgments

This work is supported by the National Science Foundation under Award No. NSF 2523284. The STM images were made available by Ezra Bussmann as the result of the work performed at the Center for Integrated Nanotechnologies, an Office of Science User Facility operated for the U.S. Department of Energy (DOE) Office of Science by Los Alamos National Laboratory (Contract 89233218CNA000001) and Sandia National Laboratories (Contract DE-NA-0003525).

**Data Availability**

The original reconstruction framework is available in Sergei V. Kalinin's repository (https://github.com/SergeiVKalinin/spm-reconstruction). The corresponding author's development repository is https://github.com/1996suryamandalreddy-maker/spm-reconstruction. The accompanying JAP_supplementary_repository snapshot contains the analysis notebooks, figure-generation code, editable PowerPoint assembly, frozen numerical results, and file manifest for this revision. Synthetic figures and quantitative panels can be regenerated from this snapshot; regeneration of experimental image components requires the original acquisitions and stored reconstruction arrays, whose access conditions must be confirmed by the authors. A versioned archival release and DOI have not yet been assigned.

**REFERENCES**

[1]G. Binnig, H. Rohrer, C. Gerber, and E. Weibel, "7 × 7 reconstruction on Si(111) resolved in real space," Phys. Rev.Lett. **50**, 120 (1983).

[2]G. Binnig, C. F. Quate, and C. Gerber, "Atomic force microscope," Phys. Rev. Lett. **56**, 930 (1986).

[3]R. García and R. Pérez, "Dynamic atomic force microscopy methods," Surf. Sci. Rep. **47**, 197 (2002).

[4]N. Jalili and K. Laxminarayana, "A review of atomic force microscopy imaging systems: application to molecular metrology and biological sciences," Mechatronics **14**, 907 (2004).

[5]D. Nečas and P. Klapetek, "Gwyddion: an open-source software for SPM data analysis," Cent. Eur. J. Phys. **10**, 181 (2012).

[6]I. Horcas, R. Fernández, J. M. Gómez-Rodríguez, J. Colchero, J. Gómez-Herrero, and A. M. Baró, "WSXM: A software for scanning probe microscopy and a tool for nanotechnology," Rev. Sci. Instrum. **78**, 013705 (2007).

[7]Y. Wu, Y. Fang, Z. Fan, and C. Liu, "Accurate morphology characterization using atomic force microscopy via vertical drift correction and illusory slope elimination," Microsc. Microanal. **27**, 1366 (2021).

[8]S.-W. W. Chen and J.-L. Pellequer, "DeStripe: frequency-based algorithm for removing stripe noises from AFM images," BMC Struct. Biol. **11**, 7 (2011).

[9]Y. Zhang, Y. Li, Z. Song, Z. Wang, J. Qian, and J. Yao, "A novel method to remove impulse noise from atomic force microscopy images based on Bayesian compressed sensing," Beilstein J. Nanotechnol. **10**, 2346 (2019).

[10]M. Li, J. Rieck, B. Noheda, J. B. T. M. Roerdink, and M. H. F. Wilkinson, "Stripe noise removal in conductive atomic force microscopy," Sci. Rep. **14**, 3931 (2024).

[11]M. Schimmack and P. Mercorelli, "An on-line orthogonal wavelet denoising algorithm for high-resolution surface scans," J. Franklin Inst. **355**, 9245 (2018).

[12]J. S. Villarrubia, "Algorithms for scanned probe microscope image simulation, surface reconstruction, and tip estimation," J. Res. Natl. Inst. Stand. Technol. **102**, 425 (1997).

[13]F. Tian, X. Qian, and J. S. Villarrubia, "Blind estimation of general tip shape in AFM imaging," Ultramicroscopy **109**, 44 (2008).

[14]D. Croft, G. Shed, and S. Devasia, "Creep, hysteresis, and vibration compensation for piezoactuators: Atomic force microscopy application," J. Dyn. Syst. Meas. Control **123**, 35 (2001).

[15]R. V. Lapshin, "Feature-oriented scanning methodology for probe microscopy and nanotechnology," Nanotechnology **15**, 1135 (2004).

[16]B. S. Salmons, D. R. Katz, and M. L. Trawick, "Correction of distortion due to thermal drift in scanning probe microscopy," Ultramicroscopy **110**, 339 (2010).

[17]M. P. Yothers, A. E. Browder, and L. A. Bumm, "Real-space post-processing correction of thermal drift and piezoelectric actuator nonlinearities in scanning tunneling microscope images," Rev. Sci. Instrum. **88**, 013708 (2017).

[18]C. Ophus, J. Ciston, and C. T. Nelson, "Correcting nonlinear drift distortion of scanning probe and scanning transmission electron microscopies from image pairs with orthogonal scan directions," Ultramicroscopy **162**, 1 (2016).

[19]T. Dickbreder, F. Sabath, L. Höltkemeier, R. Bechstein, and A. Kühnle, "unDrift: A versatile software for fast offline SPM image drift correction," Beilstein J. Nanotechnol. **14**, 1225 (2023).

[20]J. H. Kindt, J. B. Thompson, M. B. Viani, and P. K. Hansma, "Atomic force microscope detector drift compensation by correlation of similar traces acquired at different setpoints," Rev. Sci. Instrum. **73**, 2305 (2002).

[21]I. Azuri, I. Rosenhek-Goldian, N. Regev-Rudzki, G. Fantner, and S. R. Cohen, "The role of convolutional neural networks in scanning probe microscopy: a review," Beilstein J. Nanotechnol. **12**, 878 (2021).

[22]Y. Liu, Q. Sun, W. Lu, H. Wang, Y. Sun, Z. Wang, X. Lu, and K. Zeng, "General resolution enhancement method in atomic force microscopy using deep learning," Adv. Theory Simul. **2**, 1800137 (2019).

[23]Y. Wu, Y. Fang, and X. Ren, "A high-efficiency Kalman filtering imaging mode for an atomic force microscopy with hysteresis modeling and compensation," Mechatronics **50**, 69 (2018).

[24]P. Zheng, H. He, Y. Gao, P. Tang, H. Wang, J. Peng, L. Wang, C. Su, and S. Ding, "Speeding up the topography imaging of atomic force microscopy by convolutional neural network," Anal. Chem. **94**, 5041 (2022).

[25]H. Jung, G. Han, S. J. Jung, and S. W. Han, "Comparative study of deep learning algorithms for atomic force microscopy image denoising," Micron **161**, 103332 (2022).

[26]F. Joucken, J. L. Davenport, Z. Ge, E. A. Quezada-López, T. Taniguchi, K. Watanabe, J. J. Velasco, J. Lagoute, and R. A. Kaindl, "Denoising scanning tunneling microscopy images of graphene with supervised machine learning," Phys. Rev. Mater. **6**, 123802 (2022).

[27]V. Kocur, V. Hegrová, M. Patočka, J. Neuman, and A. Herout, "Correction of AFM data artifacts using a convolutional neural network trained with synthetically generated data," Ultramicroscopy **246**, 113666 (2023).

[28]Y. Matsunaga, S. Fuchigami, T. Ogane, and S. Takada, "End-to-end differentiable blind tip reconstruction for noisy atomic force microscopy images," Sci. Rep. **13**, 129 (2023).

[29]L. K. S. Bonagiri, Z. Wang, S. Zhou, and Y. Zhang, "Precise surface profiling at the nanoscale enabled by deep learning," Nano Lett. **24**, 2589 (2024).

[30]J. Park, D. Y. Cheong, G. Lee, and C. E. Han, "Deep learning-based denoising for unbiased analysis of morphology and stiffness in amyloid fibrils," Comput. Biol. Med. **184**, 109410 (2025).

[31]S. Gelman, I. Rosenhek-Goldian, N. Kampf, M. Patočka, M. Rios, M. Penedo, G. Fantner, A. Beker, S. R. Cohen, and I. Azuri, "Deep learning for enhancement of low-resolution and noisy scanning probe microscopy images," Beilstein J. Nanotechnol. **16**, 1129 (2025).

[32]M. Rashidi and R. A. Wolkow, "Autonomous scanning probe microscopy in situ tip conditioning through machine learning," ACS Nano **12**, 5185 (2018).

[33]B. Alldritt, F. Urtev, N. Oinonen, M. Aapro, J. Kannala, P. Liljeroth, and A. S. Foster, "Automated tip functionalization via machine learning in scanning probe microscopy," Comput. Phys. Commun. **273**, 108258 (2022).

[34]A. Krull, P. Hirsch, C. Rother, A. Schiffrin, and C. Krull, "Artificial-intelligence-driven scanning probe microscopy," Commun. Phys. **3**, 54 (2020).

[35]K. P. Kelley, M. Ziatdinov, L. Collins, M. A. Susner, R. K. Vasudevan, N. Balke, S. V. Kalinin, and S. Jesse, "Fast scanning probe microscopy via machine learning: Non-rectangular scans with compressed sensing and Gaussian process optimization," Small **16**, 2002878 (2020).

[36]S. V. Kalinin, M. Ziatdinov, J. Hinkle, S. Jesse, A. Ghosh, K. P. Kelley, A. R. Lupini, B. G. Sumpter, and R. K. Vasudevan, "Automated and autonomous experiments in electron and scanning probe microscopy," ACS Nano **15**, 12604 (2021).

[37]R. K. Vasudevan, K. P. Kelley, J. Hinkle, H. Funakubo, S. Jesse, S. V. Kalinin, and M. Ziatdinov, "Autonomous experiments in scanning probe microscopy and spectroscopy: Choosing where to explore polarization dynamics in ferroelectrics," ACS Nano **15**, 11253 (2021).

[38]S. V. Kalinin, E. Strelcov, A. Belianinov, S. Somnath, R. K. Vasudevan, E. J. Lingerfelt, R. K. Archibald, C. Chen, R. Proksch, N. Laanait, and S. Jesse, "Big, deep, and smart data in scanning probe microscopy," ACS Nano **10**, 9068 (2016).

[39]R. K. Vasudevan, K. P. Kelley, E. Eliseev, S. Jesse, H. Funakubo, A. Morozovska, and S. V. Kalinin, "Bayesian inference in band excitation scanning probe microscopy for optimal dynamic model selection in imaging," J. Appl. Phys. **128**, 054105 (2020).

[40]R. E. Kalman, "A new approach to linear filtering and prediction problems," J. Basic Eng. **82**, 35 (1960).

[41]R. E. Kalman and R. S. Bucy, "New results in linear filtering and prediction theory," J. Basic Eng. **83**, 95 (1961).

[42]H. E. Rauch, F. Tung, and C. T. Striebel, "Maximum likelihood estimates of linear dynamic systems," AIAA J. **3**, 1445 (1965).

[43]B. D. O. Anderson and J. B. Moore, Optimal Filtering (Prentice-Hall, Englewood Cliffs, NJ, 1979).

[44]S. Särkkä, Bayesian Filtering and Smoothing (Cambridge University Press, Cambridge, UK, 2013).

[45]R. K. Mehra, "On the identification of variances and adaptive Kalman filtering," IEEE Trans. Autom. Control **15**, 175 (1970).

[46]R. K. Mehra, "Approaches to adaptive filtering," IEEE Trans. Autom. Control **17**, 693 (1972).

[47]P. J. Huber, Robust Statistics (Wiley, New York, 1981).

[48]J.-A. Ting, E. Theodorou, and S. Schaal, "A Kalman filter for robust outlier detection," in 2007 IEEE/RSJ International Conference on Intelligent Robots and Systems (IEEE, 2007), pp. 1514–1519.

[49]G. Agamennoni, J. I. Nieto, and E. M. Nebot, "Approximate inference in state-space models with heavy-tailed noise," IEEE Trans. Signal Process. **60**, 5024 (2012).

[50]A. S. Willsky and H. L. Jones, "A generalized likelihood ratio approach to the detection and estimation of jumps in linear systems," IEEE Trans. Autom. Control **21**, 108 (1976).

[51]M. Basseville and I. V. Nikiforov, Detection of Abrupt Changes: Theory and Application (Prentice-Hall, Englewood Cliffs, NJ, 1993).

[52]J. W. Woods and C. H. Radewan, "Kalman filtering in two dimensions," IEEE Trans. Inf. Theory **23**, 473 (1977).

[53]J. Biemond, J. Rieske, and J. Gerbrands, "A fast Kalman filter for images degraded by both blur and noise," IEEE Trans. Acoust. Speech Signal Process. **31**, 1248 (1983).

[54]K. Ahnert and M. Abel, "Numerical differentiation of experimental data: local versus global methods," Comput. Phys. Commun. **177**, 764 (2007).

[55]C. J. Masreliez and R. D. Martin, "Robust Bayesian estimation for the linear model and robustifying the Kalman filter," IEEE Trans. Autom. Control **22**, 361 (1977).

[56]C. Tomasi and R. Manduchi, "Bilateral filtering for gray and color images," in Proceedings of the Sixth IEEE International Conference on Computer Vision (IEEE, Bombay, 1998), pp. 839–846.

[57]P. Perona and J. Malik, "Scale-space and edge detection using anisotropic diffusion," IEEE Trans. Pattern Anal. Mach. Intell. **12**, 629 (1990).

[58]Z. Wang, A. C. Bovik, H. R. Sheikh, and E. P. Simoncelli, "Image quality assessment: from error visibility to structural similarity," IEEE Trans. Image Process. **13**, 600 (2004).

[59]T. D. B. Jacobs, T. Junge, and L. Pastewka, "Quantitative characterization of surface topography using spectral analysis," Surf. Topogr.: Metrol. Prop. **5**, 013001 (2017).

[60]T. Gneiting and A. E. Raftery, "Strictly proper scoring rules, prediction, and estimation," J. Am. Stat. Assoc. **102**, 359 (2007).

[61]V. Kuleshov, N. Fenner, and S. Ermon, "Accurate uncertainties for deep learning using calibrated regression," in Proceedings of the 35th International Conference on Machine Learning, Proceedings of Machine Learning Research Vol. 80 (PMLR, 2018), pp. 2796–2804.

[62]L. I. Rudin, S. Osher, and E. Fatemi, "Nonlinear total variation-based noise removal algorithms," Physica D **60**, 259 (1992).

[63]A. Buades, B. Coll, and J.-M. Morel, "A non-local algorithm for image denoising," in 2005 IEEE Computer Society Conference on Computer Vision and Pattern Recognition (CVPR'05) (IEEE, 2005), Vol. 2, pp. 60–65.

[64]D. L. Donoho, "Compressed sensing," IEEE Trans. Inf. Theory **52**, 1289 (2006).

[65]C. E. Rasmussen and C. K. I. Williams, Gaussian Processes for Machine Learning (MIT Press, Cambridge, MA, 2006).

[66]S. J. Julier and J. K. Uhlmann, "Unscented filtering and nonlinear estimation," Proc. IEEE **92**, 401 (2004).

[67]D. Mukherjee, K. M. Roccapriore, A. Al-Najjar, A. Ghosh, J. D. Hinkle, A. R. Lupini, R. K. Vasudevan, S. V. Kalinin, O. S. Ovchinnikova, M. A. Ziatdinov, and N. S. Rao, "A roadmap for edge computing enabled automated multidimensional transmission electron microscopy," Microsc. Today **30**(6), 10 (2022).

[68]M. G. Kapteyn, J. V. R. Pretorius, and K. E. Willcox, "A probabilistic graphical model foundation for enabling predictive digital twins at scale," Nat. Comput. Sci. **1**, 337 (2021).

[69]B. N. Slautin, Y. Liu, H. Funakubo, R. K. Vasudevan, M. A. Ziatdinov, and S. V. Kalinin, "Bayesian conavigation: Dynamic designing of the material digital twins via active learning," ACS Nano **18**, 24898 (2024).

[70]Y. Liu, B. Slautin, I. Mercer, J.-P. Maria, and S. V. Kalinin, "From closed-loop optimization to open decision making: Coupled digital twins for predictive and autonomous microscopy," arXiv:2607.05758 (2026).

Supplementary Material

# Physics-Informed Sequential Reconstruction of Scanning Probe Microscopy Images with Calibrated Uncertainty

Surya Prakash Reddy,[1] and Sergei V. Kalinin[1]

[1]Department of Materials Science and Engineering, University of Tennessee, Knoxville, Tennessee 37996, USA

## S1. Scope, data, and reproducibility

The original reconstruction framework is available in Sergei V. Kalinin's repository (https://github.com/SergeiVKalinin/spm-reconstruction). The corresponding author's development repository is https://github.com/1996suryamandalreddy-maker/spm-reconstruction. The accompanying JAP_supplementary_repository snapshot contains the analysis notebooks, figure-generation code, editable PowerPoint assembly, frozen numerical results, and file manifest for this revision. Synthetic figures and quantitative panels can be regenerated from this snapshot; regeneration of experimental image components requires the original acquisitions and stored reconstruction arrays, whose access conditions must be confirmed by the authors. A versioned archival release and DOI have not yet been assigned.

This Supplementary Material documents the implementation choices and validation records underlying the main manuscript. The accompanying JAP_supplementary_repository folder is the reviewer-facing computational snapshot used for this revision. It contains frozen numerical results, analysis scripts, figure sources, manifest checks, and a manuscript-to-code map. A public archival release, tag, commit identifier, and DOI will be assigned before final publication; no unassigned identifier is claimed here.

Primary model: M07_S2_CT_G0_P1. Exploratory ablations: S6 higher-order state, GC edge-aware lateral regularization, and P2 dual-pass fusion. Normalized image intensities are used internally; dimensional calibration is restored only for physical metrology outputs.

## S2. Controlled model family

The model identifiers encode four axes; Table S2 lists the twelve variants implemented and the M-number by which each is labelled in the code, the result tables and the manuscript figures. S2 carries height and slow-axis rate; S6 additionally carries labeled slopes and curvatures. C0 uses constant observation reliability, and CT uses the thresholded line-quality detector. G0 has no lateral post-update regularization, and GC applies the edge-aware line smoother. P1 uses one acquisition direction and P2 applies the current approximate pass fusion. The state vectors and observation model are defined explicitly in Eqs. (S1)-(S3).

$$x_k^{S2} = [h_k, d_k]^T \tag{S1}$$

$$x_k^{S6} = [s, s_x, s_k, s_{xx}, s_{xk}, s_{kk}]^T \tag{S2}$$

$$y_k = Cx_k + v_k, \quad C = [1, 0, \dots, 0] \tag{S3}$$

Table S1. Controlled reconstruction axes.

| Axis | Level | Meaning | Interpretive status |
|---|---|---|---|
| State | S2 | State vector defined in Eq. (S1) | Primary; observable under height measurement |
| State | S6 | State vector defined in Eq. (S2) | Exploratory; only a three-state core is observable |
| Confidence | C0 | Constant observation quality | Control |
| Confidence | CT | Line wise robust feature product | Primary but scale sensitive |
| Lateral | G0 | Independent fast-axis columns | Primary |
| Lateral | GC | Edge-aware post-update smoothing | Exploratory heuristic |
| Pass | P1 | One direction | Primary |
| Pass | P2 | Quality-weighted mean; fused quality follows Eq. (S13) | Exploratory; not exact information fusion |

Every M-number that appears in the manuscript, figures, result tables, and code is mapped to its S/C/G/P combination and interpretive role. Directory names under spm_reconstruct/ carry the same encoding, so the mapping is verifiable in the repository. Four of the sixteen logical combinations are not implemented: lateral regularization (GC) is never paired with unweighted confidence (C0). The G0/GC axis labels identify lateral coupling and are unrelated to the RTS smoother gain in main-text Eq. (18); likewise, S2/S6 are state-family labels rather than the innovation variance in main-text Eq. (12).

**Table S2. Model index**

| Model | State | Confidence | Lateral | Pass | Role |
|---|---|---|---|---|---|
| M01 | S2 | C0 | G0 | P1 | Unweighted single-pass control |
| M02 | S2 | C0 | G0 | P2 | Unweighted dual-pass control (metrology bias case) |
| M07 | S2 | CT | G0 | P1 | PRIMARY model for experimental interpretation |
| M08 | S2 | CT | G0 | P2 | Dual-pass counterpart of M07; paired-input diagnostic only |
| M09 | S2 | CT | GC | P1 | Lateral-regularization ablation |
| M10 | S2 | CT | GC | P2 | Lateral + dual-pass ablation |
| M19 | S6 | C0 | G0 | P1 | S6 unweighted single-pass ablation |
| M20 | S6 | C0 | G0 | P2 | S6 unweighted dual-pass ablation |
| M25 | S6 | CT | G0 | P1 | S6 counterpart of M07 |
| M26 | S6 | CT | G0 | P2 | S6 counterpart of M08 |
| M27 | S6 | CT | GC | P1 | S6 + lateral ablation |
| M28 | S6 | CT | GC | P2 | S6 + lateral + dual-pass ablation |
| — | — | — | — | — | Not implemented: S2/S6 × C0 × GC × P1/P2 (four combinations) |

### S3. Numerical implementation

The core implementation evaluates one Kalman and Rauch–Tung–Striebel (RTS) recursion per fast-scan column. Observation quality modifies the measurement covariance according to Eq. (S4), matching main-text Eq. (8). Equation (S5) uses the same innovation-based covariance update as main-text Eq. (17): the predicted observation variance is subtracted from the mean squared innovation before applying a lower bound and exponential averaging. The default memory factor is 0.950, as listed in Table S3. The update is applied after burn-in only when mean line quality exceeds 0.7; otherwise the previous measurement variance is retained.

$$R_{eff,k} = \frac{R_k}{\max(q_k, \varepsilon)} \tag{S4}$$

$$\hat{R}_k = \alpha_R \hat{R}_{k-1} + (1-\alpha_R)\max\big(\langle \nu_k^2 \rangle_x - \boldsymbol{C}\langle \boldsymbol{P}_{k|k-1} \rangle_x \boldsymbol{C}^{\mathrm{T}}, R_{\min}\big), \tag{S5}$$

Here ν is the innovation, P is the predicted state covariance, C projects the state onto height, and angle brackets denote an average over fast-scan positions. The variance floor equals 0.1 times the baseline measurement variance. Subtracting the predicted state contribution prevents its uncertainty from being counted again as measurement noise.

Implemented values are read from spm_reconstruct/_base.py rather than inferred from prose. Process-noise standard deviations use the mathematical subscripts shown in Table S3, distinguishing them from the posterior standard deviation used for uncertainty. Line 0 is initialized directly from the first observation. In CT variants, the forward pass is run twice: first to obtain the innovation RMS used by the tip-change factor and again with the updated quality map, before one RTS backward pass.

The standard reconstruction wrapper uses a burn-in equal to the larger of 25 lines and the integer part of 10% of the scan-line count. The illustrative estimator-anatomy notebook used for main-text Figure 2 deliberately uses eight lines, a 200 × 64 step-terrace case, and seed 16; this illustration is separate from the benchmark and calibration archives.

**Table S3. Parameter inventory**

| Symbol / setting | Implemented value | Source in the code |
|---|---|---|
| $\sigma_{w,s}$ (height) | 0.020 | DEFAULTS_S2 |
| $\sigma_{w,sk}$ (slow-axis rate) | 0.010 | DEFAULTS_S2 |
| $\sigma_{w,sx}$ / $\sigma_{w,sxx}$ / $\sigma_{w,sxk}$ / $\sigma_{w,skk}$ (S6 only) | 0.015 / 0.005 / 0.005 / 0.003 | DEFAULTS_S6_EXTRA |
| $Q$ (S2) | diag($4.0 \times 10^{-4}$, $1.0 \times 10^{-4}$) | make_Q_s2 |
| $Q$ (S6) | diag($4.0 \times 10^{-4}$, $2.25 \times 10^{-4}$, $1.0 \times 10^{-4}$, $2.5 \times 10^{-5}$, $2.5 \times 10^{-5}$, $9.0 \times 10^{-6}$) | make_Q_s6 |
| $\sigma_{meas}$ | 0.050 | DEFAULTS_S2 |
| $R_0$ | $2.5 \times 10^{-3} = \sigma_{meas}^2$ | M0x model.py |
| $R_{min}$ | $2.5 \times 10^{-4} = 0.1\, R_0$ | kalman_forward |
| $\alpha_R$ | 0.950 | DEFAULTS_S2 |
| $\varepsilon$ (quality floor) | $10^{-6}$ | kalman_forward: q = max(quality, 1e-6) |
| burn-in | max(25, floor(0.10 N_k)) lines | SPMReconstructor.reconstruct |
| adaptive-R gate | Apply after burn-in when mean line quality > 0.7 | kalman_forward |
| $P_0$ | 10.0 I | make_P0_s2 / make_P0_s6 |
| State initialization | line 0 height set to the first observation | kalman_forward |
| Clipping, height (both passes) | [-0.3, 1.4] | kalman_forward / kalman_backward |

| Symbol / setting | Implemented value | Source in the code |
|---|---|---|
| Clipping, slow-axis slope (S6) | [-0.3, 0.3] | kalman_forward / kalman_backward |
| Clipping, slow-axis curvature (S6) | [-0.05, 0.05] | kalman_forward / kalman_backward |
| Covariance regularization | symmetrize + $10^{-6}$ I (forward), + $10^{-8}$ I (smoother) | kalman_forward / kalman_backward |
| CT: $\theta_c, \alpha_c$ | 0.30, 30.0 | DEFAULTS_CT |
| CT: $\theta_o, \alpha_o$ | 0.20, 25.0 | DEFAULTS_CT |
| CT: $\tau_{cons}$ | 0.25 | DEFAULTS_CT |
| CT: $\theta_t, \alpha_t$ | 3.50, 4.00 | DEFAULTS_CT |
| GC: $\beta$, uncertainty threshold | 2.5, 3.0 (x σ_meas) | DEFAULTS_GC |
| GC correlation half-width | clip(ceil(2ℓ), 3, 15); ℓ estimated per image | estimate_ell |
| P2 fusion parameters | 0.40, 0.30 | DEFAULTS_P2 |
| Factorial bootstrap | 10,000 paired-record draws, seed 81000, 2.5/97.5 percentiles | make_figure_04_factorial.py |

For numerical robustness, normalized height is clipped to [−0.3, 1.4], and selected derivative states are bounded. These safeguards prevent divergence but break an exact linear–Gaussian posterior interpretation. Heuristic lateral regularization (GC) likewise changes the point estimate without a full joint covariance update. Accordingly, the raw posterior standard deviation σ is treated as a relative diagnostic unless independently calibrated; the separate M08 calibration experiment is described in Sec. S7.

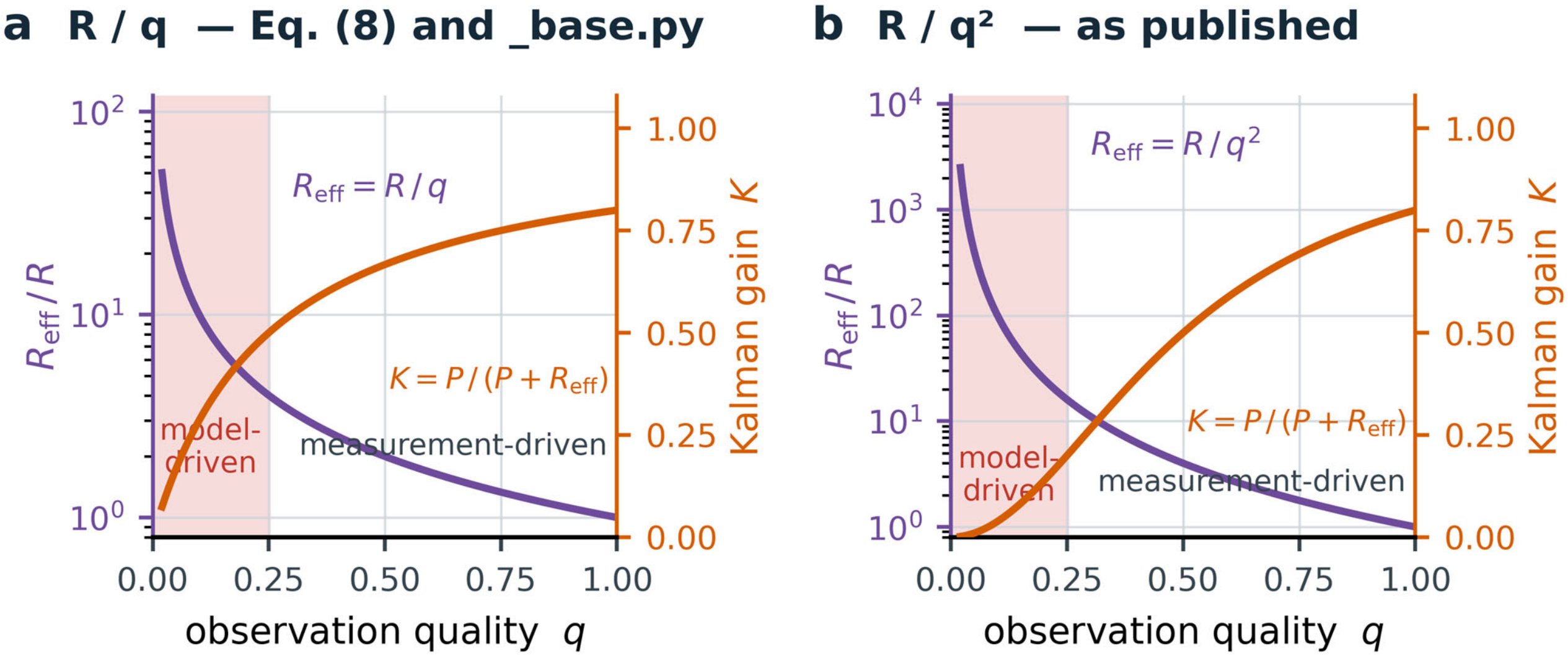

Figure S1. Observation-quality weighting. The implemented R/q mapping is compared with the more aggressive R/q squared alternative. All manuscript results and corrected schematics use R/q. The panel-b label "as published" is inherited from the earlier schematic; it denotes the alternative weighting, not the implementation used here.

### S4. Line-quality detector and scale sensitivity

For each scan line, the CT detector combines robust amplitude, excess high-frequency content, cross-line disagreement, and optional tip/pass disagreement. The implemented feature definitions and quality product are given in Eqs. (S6)-(S9). Thresholds are defined in normalized feature space, so scale sensitivity remains a limitation; the strongly periodic MOBO3 calibration scan is retained as a documented detector-failure case.

$$D_k = \text{mean}(|y_k - \text{median}(y_k)|) \tag{S6}$$

$$O_k = \text{std}(\Delta y_k) \tag{S7}$$

$$q_{c,k} = \text{sigm}[-\alpha_c(D_k - \theta_c)], \quad q_{o,k} = \text{sigm}[-\alpha_o(O_k - \theta_o)] \tag{S8}$$

$$q_k = q_{c,k} q_{o,k} q_{cons,k} q_{t,k}, \quad 0 \le q_k \le 1 \tag{S9}$$

### S5. Synthetic benchmark and factorial effects

The full archive contains 2880 model-stage records: 120 synthetic cases, twelve implemented variants, and forward/RTS outputs. Four of sixteen logical S/C/G/P combinations are not implemented: GC is paired only with CT. Figure S2 retains unnormalized metrics and runtime; main-text Figure 5 supplies a complementary within-regime radar summary. Radar scores use the extrema over all twelve variants separately for each metric and corruption regime, with error metrics reversed; these normalized scores must not be compared across spokes as absolute effects. Figure 4 and Table S4 remain the evidence for paired component effects.

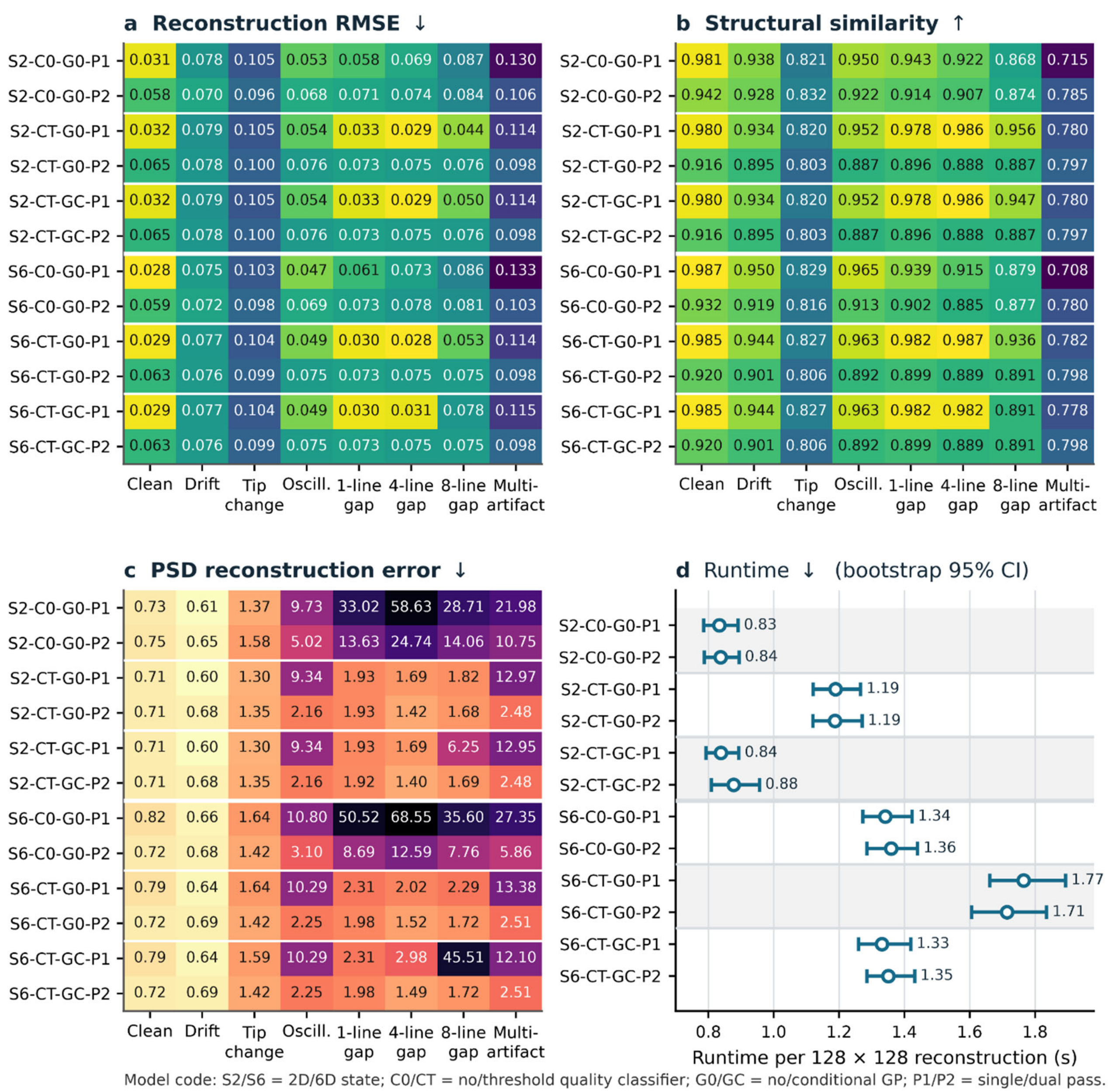

Figure S2. Model-family performance over the synthetic benchmark. Heat maps report the measured RMSE, SSIM, and power-spectral-density error by corruption regime, and the final panel reports runtime. M07 (S2-CT-G0-P1) is the prespecified primary model; lateral-regularization and dual-pass variants are interpreted as ablations rather than replacements for the primary model. The legacy GC/conditional-GP legend denotes heuristic lateral regularization, as defined in Sec. S2.

Table S4. Factorial RMSE effects.

| Component contrast | Mean RMSE effect (%) | 95% interval (%) | Interpretation |
|---|---|---|---|
| C0 to CT | +9.50 | +7.12 to +11.88 | Better overall; +27.57% for crashes |
| S2 to S6 | -0.42 | -1.60 to +0.66 | No resolved overall benefit |
| G0 to GC | -3.35 | -5.38 to -1.59 | Modest penalty overall (-8.81% on crashes); heuristic covariance |
| P1 to P2 | -66.26 | -76.04 to -56.99 | Strongly detrimental current fusion |
| Filter to RTS | +6.61 | +5.77 to +7.43 | Consistent overall benefit |

Positive effects denote reduced RMSE when the component is enabled. Intervals use 10,000 bootstrap resamples of paired records; records sharing a synthetic case are not independent cases. A separate bootstrap clustered by case gives wider intervals with the same pooled directions, including 6.25%–12.79% for quality weighting and 4.90%–8.40% for RTS smoothing, as described in main-text Sec. 3.2.

**S6. External baselines and tuning**

Total variation, non-local means, compressed sensing, and Gaussian-process reconstruction are evaluated in blind and known-mask modes. TV weight 0.002, NLM h factor 2.2, and compressed-sensing lambda 0.032 were selected on ten separately seeded tuning images by minimizing failed-region RMSE subject to clean-region RMSE damage no greater than 0.01. GP length scales were estimated per image from its autocorrelation; no benchmark case was used for tuning.

Table S5. Mean external-baseline results over six crash-gap widths.

| Method | Blind gap RMSE | Blind clean RMSE | Blind SSIM | Known-mask gap RMSE |
|---|---|---|---|---|
| Raw | 0.438 | 0.014 | 0.792 | -- |
| TV | 0.435 | 0.012 | 0.794 | 0.071 |
| NLM | 0.438 | 0.016 | 0.787 | 0.082 |
| Compressed sensing | 0.365 | 0.019 | 0.817 | 0.080 |
| Gaussian process | 0.289 | 0.024 | 0.830 | 0.091 |
| Kalman plus RTS | 0.166 | 0.036 | 0.854 | 0.086 |

### S7. Held-out uncertainty calibration

The uncertainty-calibration archive is a separate 108-case experiment and is not reused from the 120-case factorial benchmark or the six-width external-baseline benchmark. Its 54 calibration cases and 54 test cases are disjoint (calibration base seeds 91000, 92000, and 93000; test base seeds 101000, 102000, and 103000; image size 72 × 96 pixels). A monotonic empirical map is fitted without access to the test cases. The frozen coverage artifact was generated with M08, the dual-pass companion to primary M07, so it demonstrates the calibration procedure rather than establishing P2 as the preferred reconstruction. Figure S3 separates the 54-case held-out experiment from the 12 crash-gap cases used for the GP comparison. Raw σ under-covers, whereas the calibrated intervals approach nominal coverage on the held-out test set. The empirical coverage reported in the table and figure is defined in Eq. (S10).

$$\text{coverage}(p) = \frac{1}{N} \Sigma\, \mathrm{I}(|e_i| \leq z_p \sigma_i) \tag{S10}$$

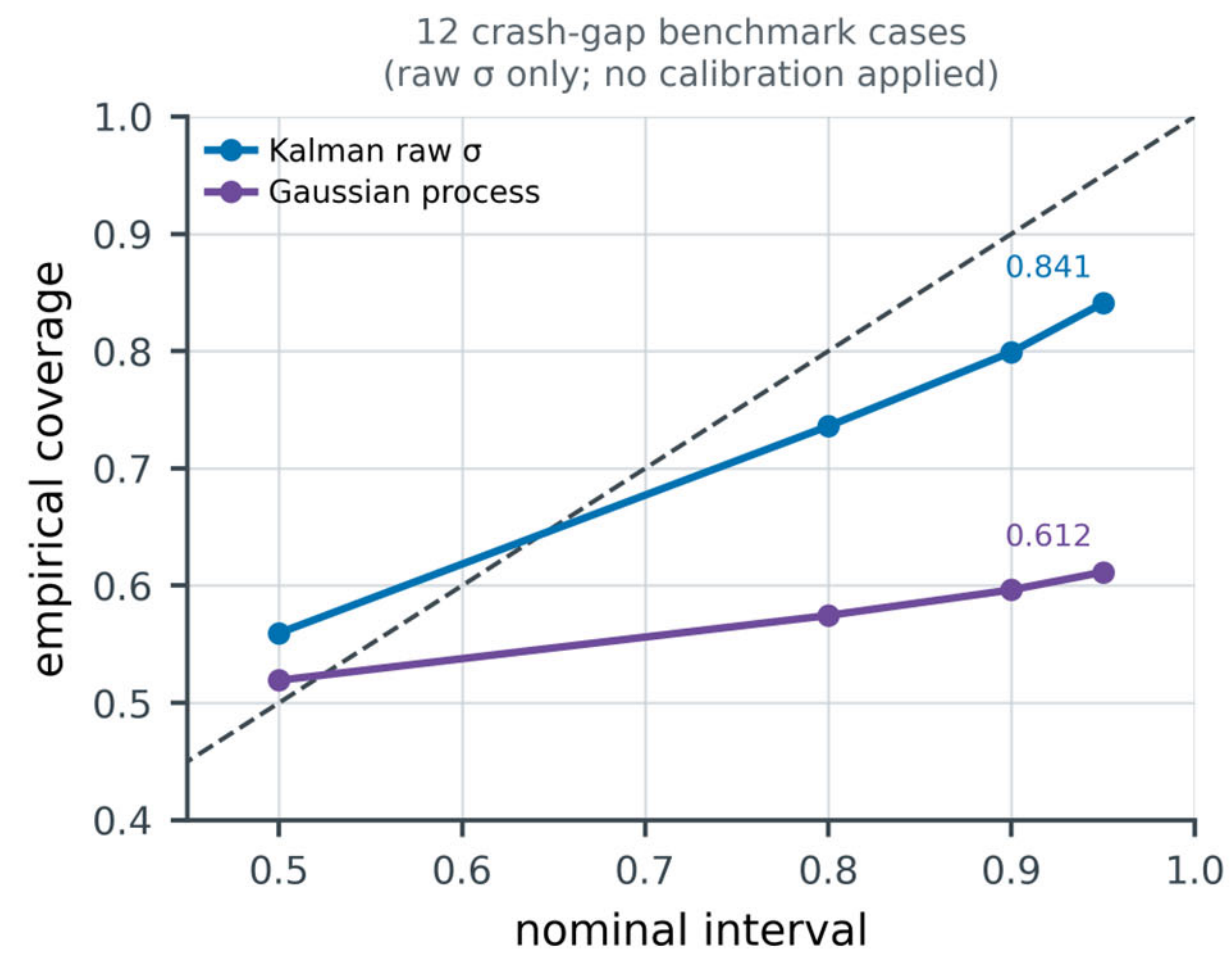

Figure S3. Empirical interval coverage by evaluation set. (a) Raw and calibrated coverage on the 54 held-out test cases from the separate uncertainty-calibration archive. (b) Uncalibrated Kalman and external Gaussian-process interval coverage on the 12 crash-gap benchmark cases. The panels are not pooled because they arise from different experiments and sample definitions.

Table S6. Held-out interval calibration.

| Quantity | All pixels | Corrupted pixels | Clean pixels |
| --- | --- | --- | --- |
| Raw nominal 95% coverage | 0.7362 | 0.7223 | 0.7362 |
| Calibrated nominal 95% coverage | 0.9191 | 0.9087 | 0.9192 |
| Calibration cases | 54 | -- | -- |
| Held-out cases | 54 | -- | -- |

## S8. Paired-pass consistency

The Red3 orientation audit is ambiguous: the selected stored orientation differs from the best data-driven candidate, and the registered correlation is approximately 0.33. Main-text Figure 7 therefore shows the stored coordinates with no additional registration or reversal and makes no agreement claim. Each pass is percentile-normalized and reconstructed independently with M07; source-only quality is displayed and reconstructed heights are restored to the source units; the grayscale montage itself does not provide a quantitative height scale. This figure should not be used to validate a registration procedure.

An opposite acquisition direction is not clean ground truth. A one-sided comparison between a corrected source pass and an unprocessed opposite pass can penalize successful artifact removal. A symmetric diagnostic applies the same reconstruction operator to both directions before comparison. The archived Figure S4 pipeline applies no subpixel registration or border mask, so its absolute values are not interchangeable with those from other orientation audits. Absolute levels remain sensitive to registration and border masking; only the contrast between comparison designs is interpreted.

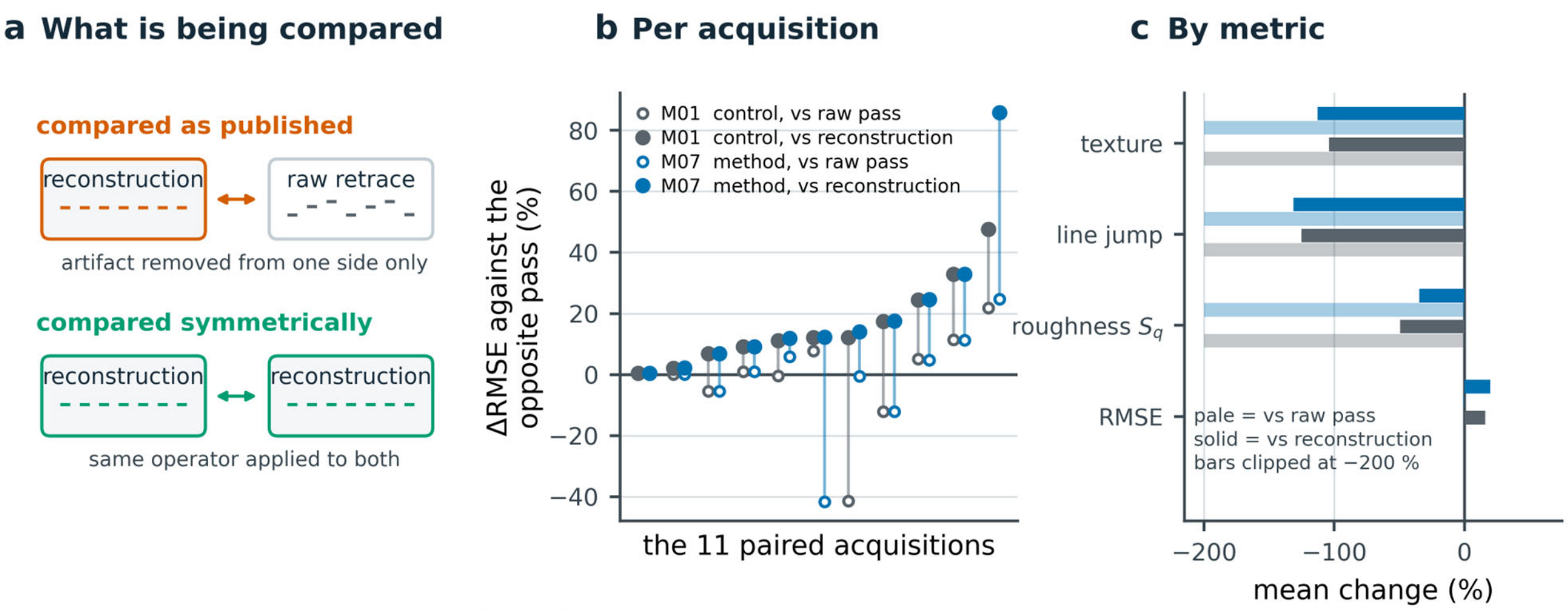

Figure S4. Effect of asymmetric and symmetric comparison designs on paired-pass consistency. Reconstruction-to-reconstruction agreement is reported as consistency, not accuracy, because neither acquisition direction provides clean ground truth.

## S9. Exploratory dual-pass diagnostics

Figure S5 illustrates the experimental paired-pass workflow using the exploratory M28 output retained in the repository. Trace and retrace are classified independently and combined by Eqs. (S11)-(S13). The quality and posterior-uncertainty maps expose line regions that receive reduced confidence. The reconstruction-minus-trace panel shows where the estimate changed; it is not a ground-truth error map. This diagnostic does not overturn the factorial result that the present dual-pass fusion rule is detrimental on average.

$$w_f = \frac{q_f}{q_f + q_b + \varepsilon} \tag{S11}$$

$$y_{fused} = w_f y_f + (1 - w_f) y_b \tag{S12}$$

$$q_{fused} = \max(q_f, q_b) \tag{S13}$$

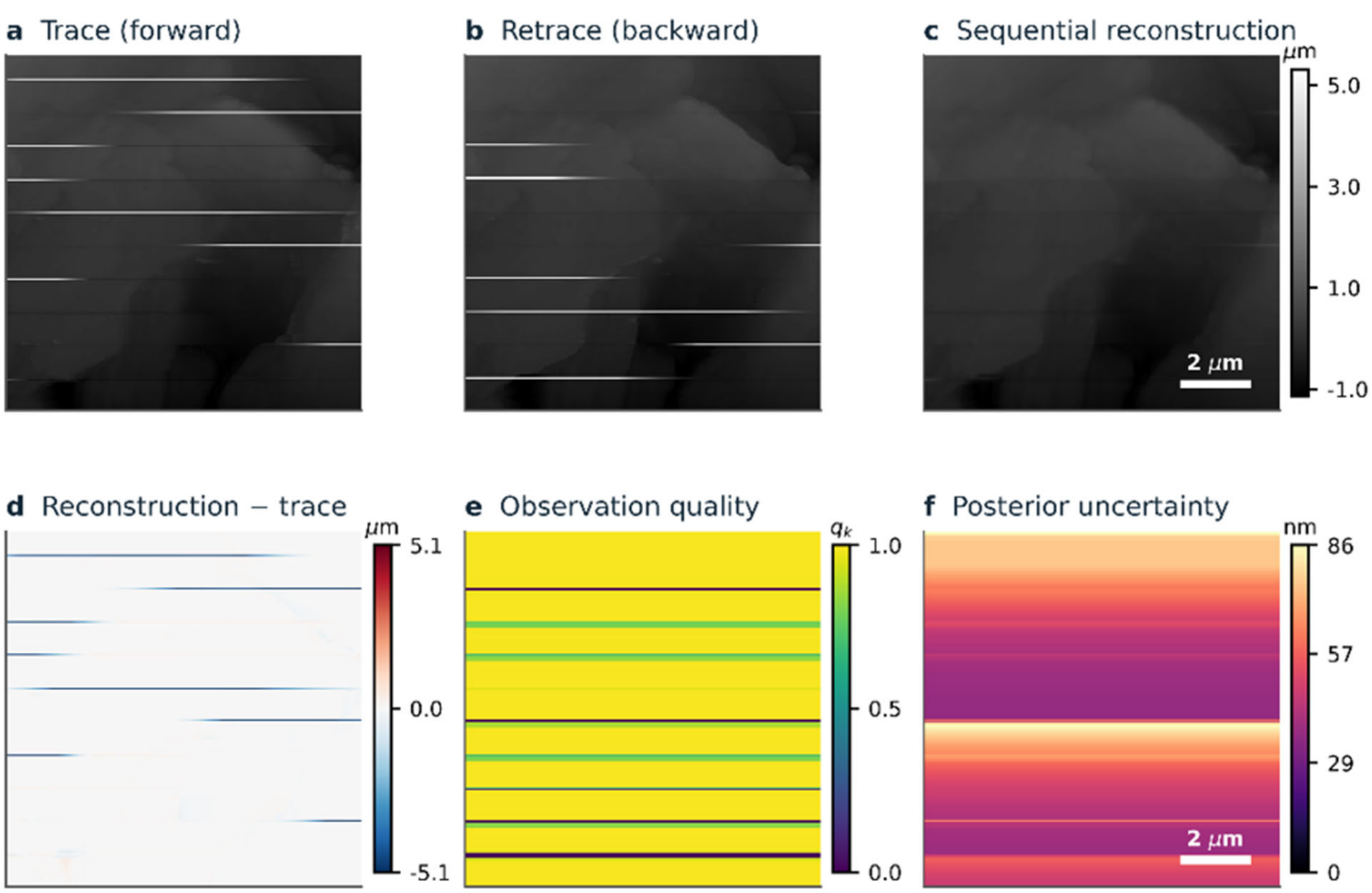

Figure S5. Experimental paired-pass reconstruction workflow for an exploratory M28 result. Trace, retrace, reconstructed height, reconstruction-minus-trace, observation quality, and relative posterior uncertainty are shown. The difference and uncertainty panels are diagnostics, not accuracy maps.

### S10. STM lattice analysis

The STM images were supplied by Azra. For Si100-Ge-3, the original metadata specify a 12 × 12 $nm^2$ field, 400 × 400 pixels, −1.9 V sample bias, and 0.1 nA tunnelling current. Ge concentration and the contributor's full name/affiliation require confirmation. Main-text Figure 8(a) shows primary M07. Panel (c) pools acquisitions and method variants within each group; its bars are medians and whiskers interquartile ranges, not the range of per-model medians quoted in the detailed numerical record. Panel (d) displays raw-image periods for Si100-Ge-1, Si111-1, Si100-Ge-2, and Si100-Ge-3; the two Au/Si acquisitions are excluded.

Dominant reciprocal-space peaks are identified in the raw STM image and evaluated at fixed coordinates after reconstruction. Lattice retention is defined by Eq. (S14). Line-artifact and high-frequency suppression are reported separately so that removal of scan-line contamination is not mistaken for improved lattice fidelity. Figure S6 shows the primary M07 reconstruction of Si100-Ge-3 together with the implemented quality and relative-uncertainty maps. Because no independent clean image exists, this is a qualitative diagnostic; quantitative claims rely on spectral retention and artifact-suppression metrics rather than visual appearance alone.

$$\rho_{lat} = \frac{A_{rec}(k^*)}{A_{raw}(k^*)} \tag{S14}$$

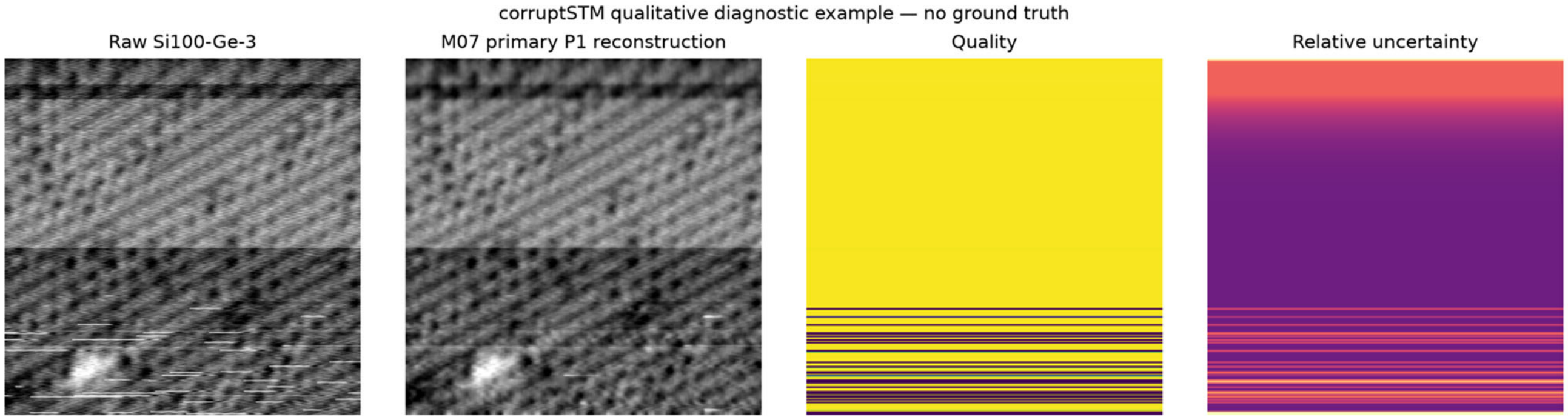

Figure S6. Qualitative STM diagnostic for Si100-Ge-3 using the primary M07 (S2-CT-G0-P1) reconstruction. The raw image, reconstruction, observation-quality map, and posterior uncertainty are shown over the same field of view. No clean ground truth is available, so the quality and uncertainty panels indicate model confidence rather than reconstruction error.

Data sets referenced in the manuscript figures are listed from results/real_data_validation/real_data_inventory.csv and analysis/results/stm_lattice_metrics.csv. Dates are inferred only from the YYMMDD prefixes of source file names. Instrument make/model, acquisition mode, setpoint, tip/sample provenance, and acquisition operator are not recorded in the supplied computational snapshot and therefore are not inferred here. Those metadata should be completed from the laboratory record before submission. Raw experimental files are available from the corresponding author subject to facility and data-use requirements.

**Table S7. Experimental data inventory**

| Data set | Source | Technique / sample | Size | Conditions | Pass | Date | Manuscript use |
|---|---|---|---|---|---|---|---|
| Red3 | 250315_Red3_MOBO.pickle | AFM Height; sample not recorded | 256 × 256; 10.0 µm | 1.0016 Hz; setpoint not recorded | Yes | 15 Mar 2025 | Main-text Fig. 7; S8 |
| MOBO1 | 250313_CaliSample_MOBO1.pickle | AFM Height; calibration grating | 256 × 256; 20.0 µm | 1.0016 Hz; setpoint not recorded | No | 13 Mar 2025 | Fig. S7; S11 |
| MOBO3 | 250313_CaliSample_MOBO3.pickle | AFM Height; calibration grating | 256 × 256; 20.0 µm | 1.0016 Hz; setpoint not recorded | Yes | 13 Mar 2025 | Fig. S7; S4, S11 |
| Cali1 | 250315_Cali1_MOBO.pickle | AFM Height; calibration grating | 256 × 256; 20.0 µm | 1.0016 Hz; setpoint not recorded | Yes | 15 Mar 2025 | Fig. S7; S11 |

| Data set | Source | Technique / sample | Size | Conditions | Pass | Date | Manuscript use |
|---|---|---|---|---|---|---|---|
| Si100-Ge-1 | STM image Si100-Ge-1 | STM; Si(100)-2×1-Ge | 300 × 300; 30.0 nm | +1.50 V; current not recorded | No | Not recorded | Main-text Fig. 8(d); S10 |
| Si100-Ge-2 | STM image Si100-Ge-2 | STM; Si(100)-2×1-Ge | 400 × 400; 20.0 nm | −1.75 V; current not recorded | No | Not recorded | Main-text Fig. 8(d); S10 |
| Si100-Ge-3 | STM image Si100-Ge-3 | STM; Si(100)-2×1-Ge | 400 × 400; 12.0 nm | −1.90 V; 0.1 nA | No | Not recorded | Main-text Fig. 8(a,d); S10 |
| Si111-1 | STM image Si111-1 | STM; Si(111) | 300 × 300; 7.5 nm | +0.75 V; current not recorded | No | Not recorded | Main-text Fig. 8(d); S10 |
| AuSi111-1 | STM image AuSi111-1 | STM; Si(111)-5×2-Au | 250 × 250; 20.0 nm | −1.90 V; current not recorded | No | Not recorded | S10; no unambiguous period |
| AuSi111-3 | STM image AuSi111-3 | STM; Si(111)-5×2-Au | 100 × 100; 9.0 nm | −1.60 V; current not recorded | No | Not recorded | S10; no unambiguous period |

### S11. AFM calibration-grating analysis

Pitch is obtained from the dominant periodicity, step height from robust separation of terrace levels, edge width from the transition region, and terrace S_q from within-level residuals. The archive contains three calibration-grating scans ($n = 3$); two scans, MOBO3 and Cali1, include retrace acquisitions. The available files identify a nominal 5.000 µm pitch and an approximately 100 nm step, but no traceable calibration certificate or certified step-height uncertainty is present in the repository. Accordingly, the analysis is reported as preservation relative to the raw/reference estimate, not as absolute vertical metrology. Most variants retain mean step height within 1.3% of that estimate, whereas the M02 dual-pass control has a +9.9 nm mean bias and a +16.5 nm worst-case bias. Using the definition in Eq. (S15), confidence weighting reduces the between-scan coefficient of variation from 4.93% to 3.69%, but this $n = 3$ repeatability result does not establish lateral calibration or absolute accuracy. The collapse to mean $q = 0.009$ on the strongly periodic MOBO3 scan is therefore retained as a detector-failure case and motivates structure-aware quality classification. The 1.3% figure is each variant's mean bias over the three scans, expressed relative to the mean raw step height; individual scans depart further, with a largest single-scan departure of 2.60% among the non-M02 variants. Both quantities are recomputed by scripts/verify_manuscript_numbers.py.

$$CV_h = 100 \times \frac{s_h}{\bar{h}} \% \quad \text{(S15)}$$

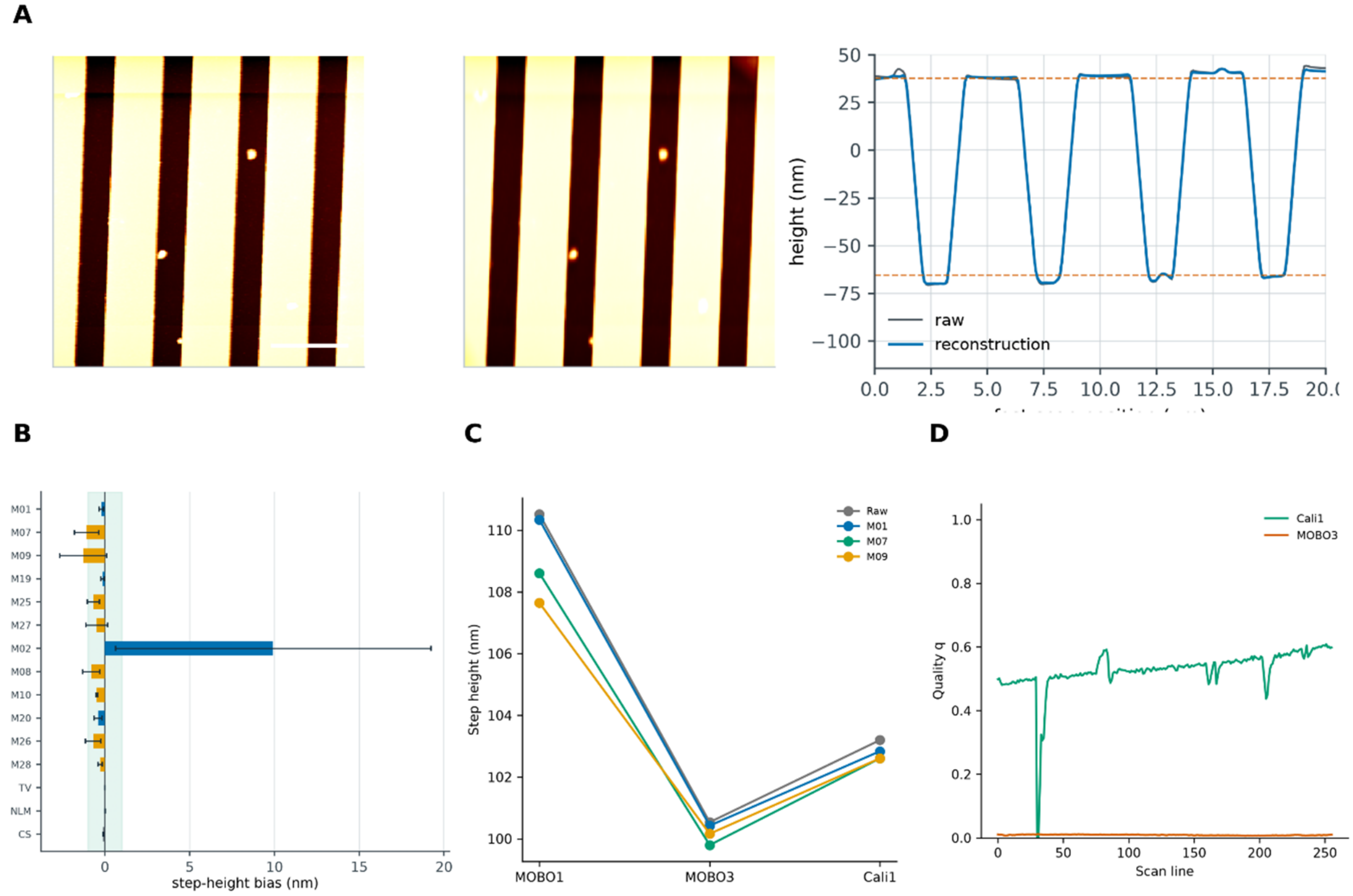

Figure S7. AFM calibration-grating validation. (A) Raw trace, archived M08 reconstruction, and mean fast-axis profiles for Cali1, a 20 µm field of a nominal 5 µm-pitch, approximately 100 nm-step grating; white image bars represent 5 µm. (B) Step-height change relative to each acquisition's own raw trace. Bars are method means and error bars are sample standard deviations across available acquisitions (three for P1; two for P2). Blue denotes unweighted Kalman variants, orange quality-weighted variants, and grey classical baselines; the shaded band is ±1 nm. The GP baseline is omitted from this panel because its mean bias is approximately +525 nm; it remains in the numerical archive. (C) Step heights across the three acquisitions for raw data, M01, M07, and M09. (D) M08 linewise quality for Cali1 and MOBO3. The latter receives persistently low quality despite a legitimate periodic structure. No certified step-height value or uncertainty is available: this is a test of preservation relative to the supplied raw readings and nominal geometry, not certificate-traceable accuracy. Sections S11 and S15 summarize the numerical metrology results; full per-acquisition values are in analysis/results/afm_calibration_metrology.csv.

## S12. Future extensions

The validated model is intentionally low order. Future work can replace the point observation with tip-shape, scanner-coordinate, hysteresis, or feedback-dynamic operators; represent persistent instrument state; and introduce sample-specific priors. Dual-pass data should be assimilated as separate likelihoods rather than through the present approximate average. These extensions are research directions and are not required for, or validated by, the results in the main manuscript.

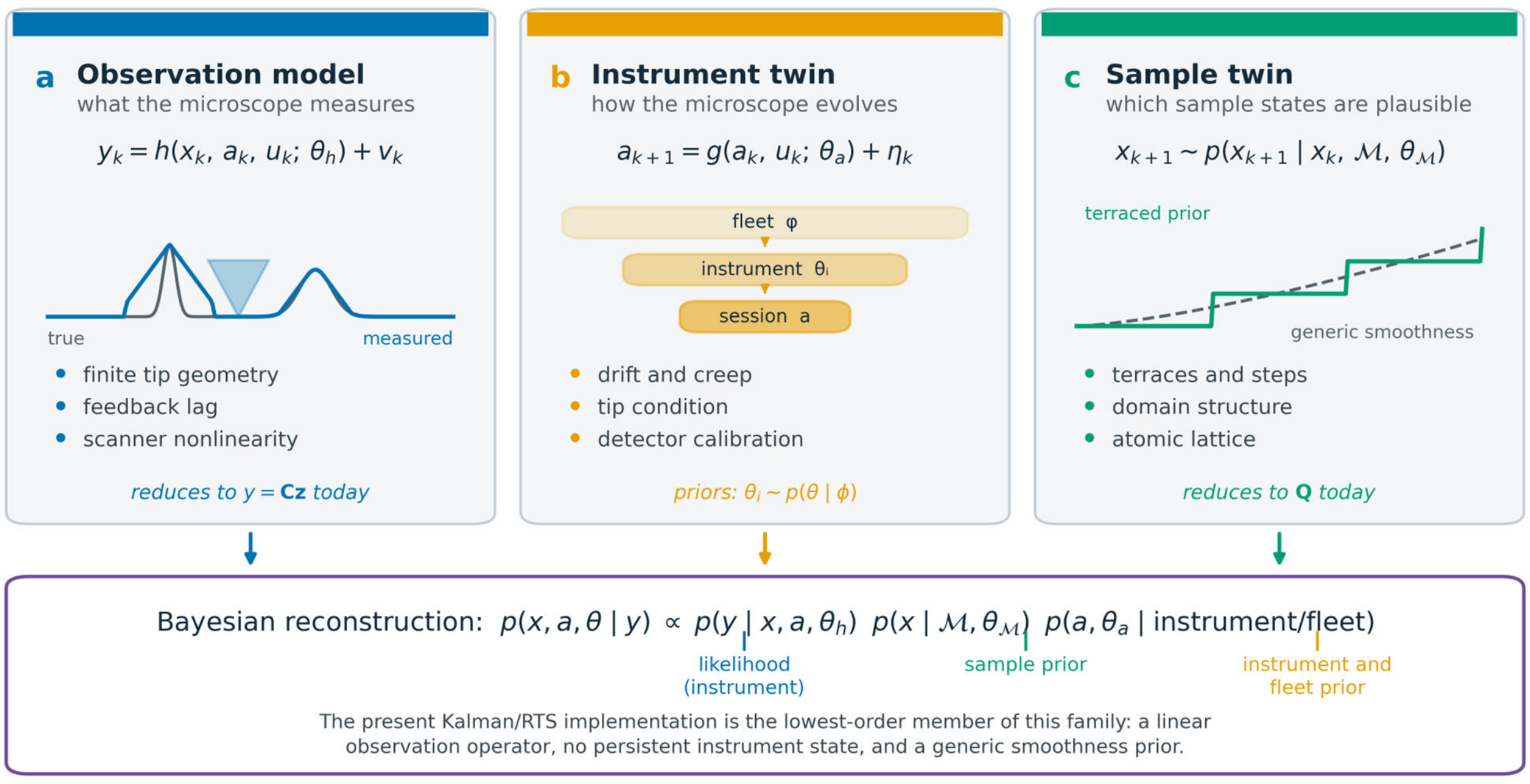

Figure S8. Conceptual future extension to coupled sample and instrument models. The diagram is prospective and does not represent a validated component of the present algorithm.

### S13. Runtime and scaling

Main-text Sec. 2.7 describes computational cost proportional to pixel count for fixed state dimension. The runtime experiment tests this scaling for the measured implementation. analysis/run_runtime_scaling.py reconstructs a deterministic crash-corrupted sinusoidal scan (seed 20260901) with the primary model M07 at six square image sizes, timing five repetitions after one untimed warm-up and reporting the median; results are written to analysis/results/runtime_scaling.csv.

Median reconstruction time is 0.104 s at 48 × 48, 0.181 s at 64 × 64, 0.409 s at 96 × 96, 0.749 s at 128 × 128, 1.454 s at 181 × 181, and 2.882 s at 256 × 256 pixels. Time per pixel remains between 43.97 and 45.69 µs per pixel across a 28-fold range in pixel count, supporting the stated linear-in-pixel-count scaling for the measured implementation and hardware. Equation (S16) states the pixel-count scaling used in this interpretation.

$$t(N) = O(N), \quad N = N_x N_y \tag{S16}$$

### S14. Computational repository and review workflow

The computational supplement contains analysis scripts, notebooks, frozen numerical tables, figure-generation sources, individual image and vector components, and an editable PowerPoint assembly. Temporary renders, local environments, duplicate manuscript drafts, and raw experimental acquisitions are excluded. Quantitative panels can be regenerated from the supplied tables; experimental image components can be reassembled without raw data, but their scientific recomputation requires the original measurements and stored reconstructions.

**Repository entry point.** README.md explains installation, the primary M07 model, scientific limitations, key commands, and the directory layout. docs/MANUSCRIPT_MAP.md links each major manuscript claim to its implementation, frozen evidence, and figure.

**Level 1 - audit without recomputation.** Reviewers can inspect results/, analysis/results/, and figures/ and run scripts/verify_frozen_results.py together with scripts/check_manifest.py. These checks validate the headline values, required output files, and distributed-file integrity.

**Level 2 - synthetic recomputation.** The deterministic generators, all twelve S/C/G/P variants, classical baselines, analysis scripts, and notebooks 07–12 are included. scripts/smoke_test.py performs a fast M07 reconstruction check; the complete factorial and uncertainty workflows require the additional runtime described in docs/REPRODUCIBILITY.md.

Level 3 — experimental recomputation. STM-lattice, AFM-calibration, and paired-pass scripts are included for transparency, but a full rerun requires original pickle/IBW files and processed per-model arrays that are not redistributed in the reviewer snapshot. Qualified reviewers may request access from the corresponding author, subject to facility, provenance, and data-use restrictions. The manuscript and repository must not imply unrestricted public redistribution until those permissions are confirmed.

**Integrity and automation.** MANIFEST.sha256 records every distributed file. The GitHub Actions workflow installs the package and runs the smoke and frozen-result checks on each push or pull request. Notebook outputs were stripped to remove machine-specific paths and avoid presenting stale execution state as a fresh rerun. Two further checks were added in this revision. scripts/verify_manuscript_numbers.py recomputes all 56 numerical values quoted in the manuscript and this Supplementary Material from the frozen archives and fails on any drift, and the snapshot was completed so that the analysis entry points import cleanly from the distributed folder alone: figure_style.py, figure_case_data.py, make_figure_02_publication.py and make_paper_figures_3_4.py are now included, the frozen-result checker points at the reorganized results/factorial_ablation/ paths, and the smoke test loads the NumPy compatibility shim so that it runs on a minimal environment.

Author action before submission. Create a public archival release of the curated JAP_supplementary_repository snapshot, record the exact Git tag and commit SHA, obtain a persistent DOI (for example through Zenodo or an institutional repository), and replace the provisional data-availability wording in the main manuscript with the final URL and access conditions. These identifiers are deliberately not fabricated in this version.

The repository does not expand the permitted scientific claims. Raw posterior σ remains a relative diagnostic until empirically calibrated; P2 and lateral coupling remain exploratory; opposite-pass comparisons measure consistency rather than accuracy; and the AFM-grating analysis establishes preservation relative to the supplied raw/reference estimate rather than certificate-traceable absolute metrology.

## S15. Detailed numerical record and figure provenance

The following numerical results have been moved out of the main narrative. Frozen CSV files retain the full precision, per-case results, model definitions, and metric-specific sample counts. These values are existing archived results, not a claim that the full benchmark was rerun for this editorial revision.

**Paired model effects**

Paired contrasts in the 2880-run factorial archive show that additional components are not uniformly beneficial (main-text Fig. 4 and Table S4). RTS smoothing reduces RMSE by 6.61% overall [95% bootstrap interval, 5.77–7.43%] and by 14.01% for crashes [12.20–15.83%]. Edge-aware lateral regularization increases RMSE by 3.35% overall [1.59–5.38%] and by 8.81% on crashes [4.25–14.32%] and does not carry a consistent posterior covariance. S6 has no resolved overall RMSE advantage over S2 (-0.42%, interval -1.60 to 0.66%). The linewise confidence detector reduces RMSE by 9.50% overall [7.12–11.88%] and by 27.57% for crashes [22.75–32.24%], although it increases RMSE modestly on the clean, drift, oscillation and tip-change strata, while the current P2 fusion increases overall RMSE by 66.26% [56.99–76.04%].

**Held-out interval results**

Evaluation includes RMSE and structural similarity (SSIM), radial power-spectral-density error, roughness bias, clean-region damage, and artifact-region error. For uncertainty, raw sigma is evaluated as a ranking diagnostic and then mapped to empirical intervals using the disjoint calibration split. On the 54 held-out test cases, pixelwise absolute error and raw sigma have Spearman correlation 0.564. On that same split, nominal 95% raw intervals cover 73.62% of all pixels and 72.23% of corrupted pixels, while calibrated intervals cover 91.91% and 90.87%, respectively. The calibrated values are close to, but do not exactly attain, nominal coverage and do not establish a fully specified Gaussian posterior.

**External-baseline results**

External baselines separate failure detection from interpolation (main-text Fig. 6). Across 30 synthetic cases spanning five surface families and six gap widths, blind Kalman+RTS reconstruction has mean failed-region RMSE 0.166 — 0.042 for a single failed line, rising monotonically to 0.467 at 32 consecutive lines, where it no longer improves on the raw data at 0.435 — compared with 0.289 for Gaussian-process, 0.365 for compressed-sensing, 0.435 for total-variation, and 0.438 for non-local-means reconstruction. Non-local means is indistinguishable from the raw data, which also scores 0.438. With the true failure mask supplied, classical methods become competitive or superior (failed-region RMSE 0.071–0.091 versus 0.086 for Kalman+RTS). The sequential method also leaves a higher absolute clean-region RMSE (0.036) than total variation (0.012); measured against the raw scans (0.014) this is a clean-region change of +0.022 for Kalman+RTS and −0.002 for total variation, so total variation slightly improves the clean regions while the sequential method perturbs them. Its blind spectral error is also comparatively high, 29.5 against 22.7 for the raw data [main-text Fig. 6(D)], although with the true mask supplied the same quantity falls to 0.48, showing that the spectral penalty comes from false detections rather than from the interpolation itself. Its advantage therefore lies mainly in detecting unreliable lines and changing their likelihood weight, with a measurable metrological trade-off.

Configuration scope: the frozen sequential comparator is M10 = (S2, CT, GC, P2). The results above must not be attributed to primary M07. The coverage archive uses M08; the benchmark GP comparison uses a separate twelve-case set.

**STM structural metrics**

Atomic-resolution STM images lack clean pixelwise ground truth, but reciprocal-space lattice peaks provide an independent physical constraint. Across six acquisitions and the six single-pass

Kalman variants, the median suppression is 2.69–2.77 times for line artifacts, 2.97–3.44 times for high-frequency content and 1.33–1.39 times for the spectral background, while 92.1–94.6% of the tracked lattice-peak amplitude is retained with no measurable peak displacement; each range is the range over the six variants of the median over the six acquisitions, which is distinct from the pooled group median and interquartile range shown in main-text Fig. 8(c). Of the four raw acquisitions with a resolvable dominant in-plane period, two admit an unambiguous assignment and agree with accepted values within 2% (0.756 nm against 0.768 nm for the Si(100)-2×1 dimer-row spacing, and 0.671 nm against 0.665 nm for the Si(111)-7×7 adatom spacing); the remaining assignments are reported without a metrological claim. Main-text Figure 8 tests preservation of physically meaningful periodic information rather than visual smoothness alone.

**AFM grating metrology**

Across three acquisitions, all Kalman variants except the unweighted P2 control retain mean step height within 1.3% of their raw reading (largest relative departure: −1.22%, M09). M02 has approximately +9.9 nm mean and +16.5 nm maximum bias across its two paired acquisitions. Raw pitches span 4.928–5.009 µm and Kalman pitches 4.927–4.992 µm. Step-height coefficients of variation are 4.93% for raw trace, 4.95% for M01, 4.34% for M07, and 3.69% for M09. These comparisons are relative to raw/reference estimates and do not establish certified accuracy. The figure's full per-acquisition values and available-method counts are in analysis/results/afm_calibration_metrology.csv.

**Figure regeneration**

The repository includes both an earlier publication assembly and the individual figure notebooks. Main-text Figure 2 uses notebooks/F02_estimator_anatomy.ipynb; its forward-filter and RTS recursions are described by main-text Eqs. (10)–(17) and (18)–(20), respectively. The earlier assembly is generated by notebooks/12_publication_revision.ipynb or analysis/publication_figures.py. Its file identifiers retain the earlier numbering: figure05 denotes the external-baseline and coverage panels now shown in main-text Figure 6, figure06 denotes the AFM paired-pass analysis now in Figure 7, and figure07 denotes the STM analysis now in Figure 8. The AFM grating analysis is supplementary Figure S7. These source identifiers must not be used as current manuscript cross-references. Add --experimental-root PATH to regenerate experimental images from the documented source layout; otherwise the assembly reuses supplied experimental image components while rebuilding quantitative panels from tables. Outputs include figure assemblies, individual panels, CSV plot data, and an assembly map. The editable PowerPoint assembly distinguishes native charts and text from microscopy images and other raster panels. Scientific recomputation of experimental panels requires the original measurements and stored reconstructions described in Sec. S14.